\documentstyle[11pt]{article}                                      %%

\expandafter\ifx\csname amssym.def\endcsname\relax \else\endinput\fi
\expandafter\edef\csname amssym.def\endcsname{%
       \catcode`\noexpand\@=\the\catcode`\@\space}
\catcode`\@=11
\def\undefine#1{\let#1\undefined}
\def\newsymbol#1#2#3#4#5{\let\next@\relax
 \ifnum#2=\@ne\let\next@\msafam@\else
 \ifnum#2=\tw@\let\next@\msbfam@\fi\fi
 \mathchardef#1="#3\next@#4#5}
\def\mathhexbox@#1#2#3{\relax
 \ifmmode\mathpalette{}{\m@th\mathchar"#1#2#3}%
 \else\leavevmode\hbox{$\m@th\mathchar"#1#2#3$}\fi}
\def\hexnumber@#1{\ifcase#1 0\or 1\or 2\or 3\or 4\or 5\or 6\or 7\or 8\or
 9\or A\or B\or C\or D\or E\or F\fi}

\font\tenmsa=msam10
\font\sevenmsa=msam7
\font\fivemsa=msam5
\newfam\msafam
\textfont\msafam=\tenmsa
\scriptfont\msafam=\sevenmsa
\scriptscriptfont\msafam=\fivemsa
\edef\msafam@{\hexnumber@\msafam}
\mathchardef\dabar@"0\msafam@39
\def\dashrightarrow{\mathrel{\dabar@\dabar@\mathchar"0\msafam@4B}}
\def\dashleftarrow{\mathrel{\mathchar"0\msafam@4C\dabar@\dabar@}}

\def\ulcorner{\delimiter"4\msafam@70\msafam@70 }
\def\urcorner{\delimiter"5\msafam@71\msafam@71 }
\def\llcorner{\delimiter"4\msafam@78\msafam@78 }
\def\lrcorner{\delimiter"5\msafam@79\msafam@79 }
\def\yen{{\mathhexbox@\msafam@55}}
\def\checkmark{{\mathhexbox@\msafam@58}}
\def\circledR{{\mathhexbox@\msafam@72}}
\def\maltese{{\mathhexbox@\msafam@7A}}

\font\tenmsb=msbm10
\font\sevenmsb=msbm7
\font\fivemsb=msbm5
\newfam\msbfam
\textfont\msbfam=\tenmsb
\scriptfont\msbfam=\sevenmsb
\scriptscriptfont\msbfam=\fivemsb
\edef\msbfam@{\hexnumber@\msbfam}
\def\Bbb#1{{\fam\msbfam\relax#1}}
\def\widehat#1{\setbox\z@\hbox{$\m@th#1$}%
 \ifdim\wd\z@>\tw@ em\mathaccent"0\msbfam@5B{#1}%
 \else\mathaccent"0362{#1}\fi}
\def\widetilde#1{\setbox\z@\hbox{$\m@th#1$}%
 \ifdim\wd\z@>\tw@ em\mathaccent"0\msbfam@5D{#1}%
 \else\mathaccent"0365{#1}\fi}
\font\teneufm=eufm10
\font\seveneufm=eufm7
\font\fiveeufm=eufm5
\newfam\eufmfam
\textfont\eufmfam=\teneufm
\scriptfont\eufmfam=\seveneufm
\scriptscriptfont\eufmfam=\fiveeufm
\def\frak#1{{\fam\eufmfam\relax#1}}

\csname amssym.def\endcsname

\makeatletter
\def\section{\@startsection {section}{1}{\z@}{-3.5ex plus -1ex minus
 -.2ex}{2.3ex plus .2ex}{\large\bf}}
\def\subsection{\@startsection{subsection}{2}{\z@}{-3.25ex plus -1ex minus
 -.2ex}{1.5ex plus .2ex}{\normalsize\bf}}
\makeatother

\makeatletter
\@addtoreset{equation}{section}

\makeatother

\newcommand{\nc}{\newcommand}
\newcommand{\rnc}{\renewcommand}

\nc{\be}{\begin{equation}}
\nc{\ee}{\end{equation}}
\nc{\bea}{\begin{eqnarray}}
\nc{\eea}{\end{eqnarray}}
\rnc{\a}{\alpha}
\nc{\ab}{\bar{\a}}
\nc{\ap}{\a^{+}}
\nc{\abm}{\ab^{-}}
\rnc{\b}{\beta}
\nc{\bb}{\bar{\b}}
\nc{\bbp}{\bb_{\zb}^{+}}
\nc{\bm}{\b_{z}^{-}}
\nc{\oa}{\overline{\a}}
\nc{\ob}{\overline{\b}}
\rnc{\gg}{\gamma}
\rnc{\d}{\delta}
\nc{\f}{\phi}
\nc{\fb}{\bar{\phi}}
\nc{\vf}{\varphi}
\nc{\p}{\psi}

\rnc{\c}{\chi}
\nc{\la}{\lambda}
\nc{\m}{\mu}
\nc{\n}{\nu}
\rnc{\o}{\omega}
\nc{\Om}{\Omega}
\rnc{\t}{\theta}
\nc{\eps}{\epsilon}
\rnc{\S}{\Sigma}
\nc{\F}{\Phi}

\nc{\trac}[2]{{\textstyle\frac{#1}{#2}}}

\nc{\ex}[1]{\mbox{e}^{\,\textstyle#1}}

\nc{\mat}[4]{\left(\begin{array}{cc}#1&#2\\#3&#4\end{array}\right)}

\nc{\som}[9]{\left(\begin{array}{ccc}#1&#2&#3\\#4&#5&#6\\#7&#8&#9%
\end{array}\right)}

\nc{\tr}{\mathop{\mbox{tr}}\nolimits}
\nc{\ad}{\mathop{\mbox{ad}}\nolimits}
\nc{\Tr}{\mathop{\mbox{Tr}}\nolimits}
\nc{\Det}{\mathop{\mbox{Det}}\nolimits}
\nc{\rk}{\mathop{\mbox{rk}}\nolimits}
\nc{\ra}{\rightarrow}
\nc{\Ra}{\Rightarrow}
\nc{\LRa}{\Leftrightarrow}
\nc{\ot}{\otimes}
\rnc{\ss}{\subset}
\nc{\nul}{\noindent\underline}
\nc{\non}{\nonumber\\}

\nc{\subs}[1]{{\vspace*{0.5cm}}%
{\noindent\underline{#1}}{\addcontentsline{toc}{subsection}{#1}}%
{\vspace*{0.3cm}}}
\nc{\zb}{\bar{z}}
\rnc{\lg}{\frak{g}}
\nc{\lt}{\frak{t}}
\nc{\lk}{\frak{k}}
\nc{\lh}{\frak{h}}
\nc{\pik}{\Pi_{\lk}}
\nc{\pip}{\Pi_{+}}
\nc{\pim}{\Pi_{-}}
\nc{\pih}{\Pi_{\lh}}
\nc{\jz}{J_{z}}
\nc{\jzh}{\jz^{\lh}}
\nc{\jzp}{\jz^{+}}
\nc{\jzm}{\jz^{-}}
\nc{\del}{\partial}
\nc{\dz}{\del_{z}}
\nc{\dzb}{\del_{\bar{z}}}
\nc{\az}{A_{z}}
\nc{\azb}{A_{\bar{z}}}
\nc{\g}{g^{-1}}
\nc{\dw}{\Delta_{W}}
\nc{\Ad}{{\mbox{Ad}}}
\nc{\ks}{Ka\-za\-ma-\-Su\-zu\-ki}
\nc{\KS}{\ks}
\nc{\ksm}{\ks\ model}
\rnc{\AA}{{\Bbb A}}
\nc{\BB}{{\Bbb B}}
\nc{\CC}{{\Bbb C}}
\nc{\PP}{{\Bbb P}}
\nc{\cpm}{\CC\PP(m)}
\nc{\cpn}{\CC\PP(n)}
\nc{\cp}[1]{\CC\PP(#1)}
\nc{\gmn}{G(m,m+n)}
\nc{\gmnk}{\gmn_{k}}
\nc{\cO}{{\cal O}}
\nc{\bcO}{\bar{\cO}}
\nc{\bO}{\bar{O}}
\nc{\oQ}{\overline{Q}}
\begin{document}
\global\parskip=4pt

%%%%%%%%% title page %%%%%%%%%%%%%%%%%%%%%%%%%%%%%%%%%%%%%%%%
\makeatletter
\begin{titlepage}
\begin{center}
%\vskip .5in
{\LARGE\bf Some General Aspects of Coset Models\\[2mm]
 and Topological Kazama-Suzuki Models}\\
\vskip .2in
{\bf Matthias Blau}\footnote{e-mail: mblau@enslapp.ens-lyon.fr; supported
by EC Human Capital and Mobility Grant ERB-CHB-GCT-93-0252}
\vskip .10in
Laboratoire de Physique Th\'eorique
{\sc enslapp}\footnote{URA 14-36 du CNRS,
associ\'ee \`a l'E.N.S. de Lyon,
et \`a l'Universit\'e de Savoie}\\
ENSLyon,
46 All\'ee d'Italie,\\
{}F-69364 Lyon CEDEX 07, France\\
\vskip .20in
{\bf Faheem Hussain}\footnote{e-mail: hussainf@ictp.trieste.it}
and {\bf George Thompson}\footnote{e-mail: thompson@ictp.trieste.it}\\
%\vskip .10in
ICTP\\
P.O. Box 586\\
34014 Trieste, Italy
\end{center}
%\begin{abstract}
\begin{small}
\noindent
We study global aspects of $N=2$ Kazama-Suzuki coset models by investigating
topological $G/H$ Kazama-Suzuki models in a Lagrangian framework based on
gauged Wess-Zumino-Witten models. To that end, we first generalize
Witten's analysis of the holomorphic factorization of bosonic $G/H$ models
to models with $N=1$ and $N=2$ supersymmetry, in the course of which we
also find some new anomaly-free and supersymmetric models based on
non-diagonal embeddings of the gauge group. We then explain
the basic properties (action, symmetries, metric independence, ...) of the
topologically twisted $G/H$ Kazama-Suzuki models. As non-trivial gauge
bundles unavoidably occur, we explain how all of the above generalizes
to that case.

We employ the path integral methods of localization and abelianization
(shown to be valid also for non-trivial bundles)
to establish that the twisted $G/H$ models can be localized to bosonic
$H/H$ models (with certain quantum corrections), and can hence be
reduced to an Abelian bosonic $T/T$ model, $T$ a maximal torus of $H$.
We also present the action and the symmetries of the coupling of these
models to topological gravity.  We determine the bosonic
observables for all the models based on classical flag manifolds and
the bosonic observables and their fermionic descendants for models
based on complex Grassmannians.  These results will be used in
subsequent publications to calculate explicitly the  chiral primary
ring of Kazama-Suzuki models.
\end{small}
%\end{abstract}
%\vfill
IC/95/340 \hfill hep-th/9510187 \hfill
{\small E}N{\large S}{\Large L}{\large A}P{\small P}-L-556/95
\end{titlepage}
\makeatother
%%%%%%%%% end of title page %%%%%%%%%%%%%%%%%%%%%%%%%%%%%%

\begin{small}
\tableofcontents
\end{small}

\setcounter{footnote}{0}

\section{Introduction}

In this article we investigate some global aspects of
Wess-Zumino-Witten (WZW) coset models in general, and of $N=2$
Kazama-Suzuki models \cite{ks} in particular.  WZW models \cite{ewwzw,ewhf}
are the basic building blocks of (rational) conformal field theories
and are, in principle,
exactly solvable due to their large symmetry algebras. As such they
form one of the cornerstones of our present understanding of string
theory.  On the other hand, world-sheet theories with $N=2$
superconformal symmetry not only display a very rich algebraic structure
\cite{lvw} but are also phenomenologically attractive as they lead
to $N=1$ space-time supersymmetry.

Of particular interest, therefore, is a class of world-sheet theories,
known as Kazama-Suzuki models, combining these two features. This large
class of $N=2$ theories was discovered by
Kazama and Suzuki \cite{ks} by determining under which
conditions the $N=1$ super-GKO coset construction \cite{gko} for a $G/H$
coset conformal field theory actually gives
rise to an extended $N=2$ superconformal symmetry. For a complete
classification, see \cite{schweigert}.

Subsequent investigations of these models, using the powerful
operator formalism of
(super-)conformal field theories \cite{lvw,flmw,lw,dg1,dg2,ht}, have
brought to light the rich and beautiful structure underlying these models,
arising from the interplay between the geometry and topology of coset
spaces on the one hand and the $N=2$ algebra on the other. Among the most
striking results is the emergence of global (topological) structures
(like the relation of the chiral primary ring \cite{lvw} to cohomology rings)
from these essentially local considerations.

In such a situation it is desirable to understand these results in a
different way, perhaps from an inherently more global point of view which
might be capable of shedding some light on the raisons d'\^etre of these
features of a field theory. Now,
certain global aspects of (two-dimensional) field theories are
frequently easier to extract from an action-based path integral formulation
of the theory. The loss of flexibility in working with a fixed action is
then compensated by the ability to employ other powerful methods of
(gauge) field theories like the path integral formalism which are quite
complementary to the standard conformal field theory techniques.

Coset models were first considered by Halpern et al \cite{halpern}.
The Lagrangian approach to coset WZW models was pioneered by Gawedzki and
Kupiainen \cite{gkcoset} and Karabali and Schnitzer \cite{kscoset} in
the bosonic case and by Schnitzer \cite{schnitzer} for the $N=1$ models.
Based on this, a Lagrangian realization of \ks\ models was then provided
in \cite{ewcp,nak}. There it was shown that, under the conditions for
$N=2$ superconformal symmetry determined in \cite{ks}, also the
Lagrangian realization of the $N=1$ coset models
as supersymmetric gauged WZW models \cite{schnitzer}
actually possesses the expected $N=2$ superconformal symmetry.

Some aspects of this Lagrangian realization
were investigated further by Nakatsu and Sugawara \cite{ns}
and Henningson \cite{heneg,henmir}. In particular, in \cite{ns}
the relation between the gauge-theoretic and the more standard
conformal field theory realization of these models was clarified
by showing how systematic use of the `complex gauge' trick
\cite{gkcoset,kscoset} leads to a free-field representation of \ks\ models.

Our starting point for this and the subsequent paper \cite{bhtgr} is
the work of Witten \cite{ewcp} who analyzed in detail the topologically
twisted $G/H =CP(1)$ models (a.k.a.\ $N=2$ minimal models) and their
coupling to topological gravity. In fact, our main aim in this paper,
which deals with the more formal and general aspects of \ks\ models,
will be to (partially) generalize the analysis of Witten to arbitrary $G/H$
\ks\ models. The results will then be used in \cite{bhtgr} and subsequent
publications to perform explicit calculations in these models.

In particular, we will establish here that a topological
$G/H$ \ks\ model can be reduced (localized)
to a bosonic Abelian topological field
theory, and we present the coupling of an arbitrary $G/H$ model to
topological gravity. The former simplifies significantly the task
of performing calculations in these models. The latter
provides us with a Lagrangian
realization for a large class of topological matter-gravity systems
and hence topological string theories and a corresponding generalization
of the intersection theory studied in \cite{ewcp}.

While the main subject of this paper are the twisted $N=2$ models, it is
natural to proceed to them via the original bosonic cosets and then their
supersymmetric extensions.
In order to gain a good understanding of the topological \ks\ models,
which are not quite topological field theories of the (well understood)
cohomological kind,
we find it useful to start with the approach to (bosonic) coset models
suggested by Witten in \cite{ewhf}, based on wave functions constructed
from anomalously gauged WZW models. We first generalize this analysis
to models with $N=1$ and $N=2$ supersymmetry.  While this
generalization is quite straightforward, at least formally\footnote{Our
discussion in that section will be at a level where we do not
regularize explicitly quadratic operators - the required quantum
corrections are, however, known to lead to the familiar shifts in the
level $k$.} and in principle, it sheds some light on some of
the issues we will discuss in the more specific context of topological
\ks\ models later on (the supersymmetries, the metric (in-)dependence,
the coupling to topological gravity). As a by-product it also provides
us with plenty of new anomaly-free supersymmetric (and topological)
models arising from non-diagonal embeddings of the gauge group. We
will, however, postpone a more detailed discussion of these new models
and just sketch their construction here.

The other ingredient in our analysis of the topological \ks\ models
are various localization techniques for functional integrals
(see \cite{btlocrev} for a review). A combination of the
localization techniques used e.g.\ in \cite{ewcp} and \cite{btloc}
with the method of diagonalization introduced in \cite{btver} (and
analysed further in \cite{ewgr,btdia} - see also \cite{hjs} for
very nice recent applications of this technique) will allow us to simplify
the original theory to the extent that explicit calculation of
correlation functions becomes feasible. In fact, we will be able to reduce
the original non-Abelian, non-linear and supersymmetric topological $G/H$
\ks\ models to a bosonic Abelian $T_{H}/T_{H}$ model ($T_{H}$ a maximal
torus of $H$) - with certain quantum corrections which in this section
we will take care to keep track of.

As the partition function of the topological \ks\ models generically
vanishes because of ghost number anomalies, to make the above result
meaningful and useful we also determine the (classical) observables
in a number of models (complex Grassmannians, flag manifolds $G/T_{G}$
of classical groups $G$). In all these models we find that the competing
demands of gauge invariance and supersymmetry can be met and
yield $r$ (the rank of $G$) independent generators of the classical
algebra of observables.

In \cite{bhtgr}, we will then use these result to get a fairly
explicit expression for correlators in topological \ks\ models based on
complex Grassmannians, and to establish the relation (mentioned above)
between the ring of topological observables and the classical cohomology
ring of the corresponding Grassmannian. What is interesting from the
present point of view of gauged WZW models is that, while one is gauging
the adjoint action of $H$ on $G$, the cohomology one finds is not that
of the (rather singular) space $G/\mbox{Ad}H$ but rather that of
the right (or left) coset $G/H$. We still have no completely satisfactory
explanation for how this comes about, save in a particular topological
sector in which essentially only the fermionic part of the theory plays
a role which is more obviously related to the ordinary coset $G/H$ (or,
rather, its tangent bundle) \cite{bhtgr}.

We should warn the reader that through most of
the paper we will
make use of the usual gauged Wess-Zumino action. That is, we will treat
all the fields as appropriately valued maps. This does not really
suffice when one is working with non-trivial bundles, as we are forced
to, for then the fields are really sections. However, we
proceed as if the fields are maps and return, at the end, to the
subtleties involved in defining the gauged Wess-Zumino-Witten term for
non-trivial bundles. We will see that our conclusions based, as they
were, on the naive considerations at the start
of the paper are indeed correct.

More systematically, this paper is organized as follows. In section 2
we review the reults of \cite{ewhf} on the holomorphic factorization
of the partition function of bosonic coset models and generalize them
in two ways, extending the analysis to correlation functions and
$N=1$ supersymmetric coset models. We also sketch a construction of
some new anomaly free supersymmetric models arising in this way.

In section 3, we deal with the models of primary interest in this paper,
namely $N=2$ coset models and their topological partners. We recall the
Lagrangian realization of the $N=2$ and topological \ks\ models and
their basic algebraic and symmetry properties. We also establish the
metric independence and holomorphic factorization of the twisted models
from the wave function point of view.

In section 4, we discuss the localization of the topological $G/H$ \ks\
model to a bosonic $H/H$ model (generalizing the corresponding result
for $SU(2)/U(1)$ in \cite{ewcp}, albeit using a slightly different
method), and the subsequent abelianization to a $T_{H}/T_{H}$ model
(following \cite{btver}).

In section 5, we first of all give a direct proof that the model we are
dealing with is a topological field theory by showing
that the energy-momentum tensor of the twisted \ks\ model
is BRST exact modulo the equations of motion of the gauge field. We
then present the coupling of the topological \ks\ model to topological
gravity. This coupling turns out to be the more or less obvious
generalization of Witten's result for $SU(2)/U(1)$ \cite{ewcp}
in the case that $G/H$ is hermitian symmetric. In general, however,
one finds that certain additional terms are also present in the Lagrangian.

In section 6 we analyze the bosonic observables
in the classes of models mentioned above, as well as their fermionic
descendants (pointing out the complications that arise when applying the
usual descent procedure, familiar from cohomological field theories,
in the somewhat different present setting).

Finally, in sections 7 and 8 we return to the issue
of what the
definition of the theory should be when the bundles in question are
non-trivial. The definition that we take is adapted from that of
\cite{gaw} and \cite{hori}. The main
technical result that we obtain is an appropriate generalization of
the Gawedzki-Hori definition to anomously gauged WZW models, i.e.
wavefunctions. On the basis of that result it is then practically
guaranteed that the analysis performed in
sections 2 through 6 remain valid in the general setting and we verify
this explicitly. Nevertheless, as pointed out by Hori \cite{hori}, taking
account of the non-trivial topological sectors is important if one is
to resolve the `fixed point' problem \cite{fp} that arises in the
context of conformal field theory.

As should be clear from the above description of what we do and what we
do not do in this paper and in \cite{bhtgr}, this is but a first step
towards a full understanding of the global properties of \ks\ models.

For example, one would like to have a better {\em a priori} understanding
from the present point of view of
what it is that one is calculating in these models (namely, cohomology
rings of homogeneous spaces and generalizations thereof
\cite{lvw,dg2,bhtgr}).

Clearly, also the topological matter
systems coupled to topological gravity we construct in section 5 are
worthy of further study, in particular, their interpretation in terms
of intersection theory on (a suitable cover of) the moduli space of
curves (as in \cite{ewcp}). Also the description of these models
in terms of integrable hierarchies
associated with $G/H$ (presumably related to the hierarchies studied in
\cite{lerche}) needs to be elucidated.

\section{Factorization of Bosonic and $N=1$ Coset Models}

We begin with a brief review of the salient features of gauged
Wess-Zumino-Witten (WZW) models. Sections 2.1, 2.2 and
2.4 are essentially a review of the pertinent results of \cite{ewhf} while
the other parts of this section contain generalizations thereof
(primarily to supersymmetric models).

The action of the
WZW model at level $k$ for a compact semi-simple group $G$ is
$kS(g)$ where
\be
S(g) = -\frac{1}{8\pi}\int_{\S}g^{-1}dg * g^{-1}dg -
i \Gamma(g) , \label{WZW}
\ee
and
\be
\Gamma(g) = \frac{1}{12\pi}\int_{M}(g^{-1}dg)^{3}
\ee
with $\partial M = \S$ and $g$ a map form the two-dimensional closed
surface $\S$ to the group $G$ ($g$ also denotes its extension to $M$).
Here and in the following a trace $\Tr$ (Killing-Cartan form of $G$)
will always be understood in integrals of Lie algebra valued forms.
We also assume that $\Tr$ has been normalized in such a way that the
quantum theory is well-defined (independent of the extension of $g$ to
$M$) for integer $k$.

The action $S(g)$ has a  global $G_{L} \times G_{R}$ invariance, $g\ra
h^{-1}gl$.
It is well known, however, that one can not gauge any subgroup of
$G_{L}\times G_{R}$, due to anomalies. The condition on the subgroup
$F$ such that it
be anomaly free is as follows. Let $\Tr_{L}$ and $\Tr_{R}$ be the traces
on Lie$\,G_{L}$ and Lie$\, G_{R}$. Then, if for all $t, t' \in$ Lie$\,
F$
\be
\Tr_{L} tt' -\Tr_{R}tt' =0  \label{anomfree}\;\;,
\ee
the subgroup $F$ is anomaly free. The standard example is when $F = H
\times  H$ (acting diagonally as $g\ra h^{-1}gh$) and the resulting theory
is known as {\em the} $G/H$ model.

\subsection{Gauged WZW Models}

Witten \cite{ewhf} has shown that the WZW action gauged
by anomalous subgroups is worthy of further study. One may use these
non-gauge invariant models to establish the holomorphic factorization of
the path integral of anomaly free gauged WZW models.

Let $F = H_{L}\times H_{R} \subseteq G_{L} \times G_{R}$ be a, possibly
anomalous, subgroup. Denote the Lie algebra of $H_{L} \times H_{R}$ by
$\lh_{L} \oplus \lh_{R}$. The field $g$ that
enters in the gauged model is correctly thought of as a section of a
$H_{L} \times H_{R} $ bundle $X \rightarrow \S$. Let $(A, B)$ be a
$(H_{L} \times H_{R})$ connection on $X$.

The action we wish to consider is
\bea
S(A,B;g)&=& -\frac{1}{8\pi}\int_{\S}g^{-1}d_{{\cal A}}g * g^{-1}d_{ {\cal
A}}g  -
i \Gamma(g) \nonumber \\
& & \, \; \; + \frac{i}{4\pi}\int_{\S} \left(Adg g^{-1} + B g^{-1}dg
 \right)
 + \frac{i}{4\pi}\int_{\S} Bg^{-1}Ag
\eea
where the covariant derivative is defined by
\be
d_{ {\cal A} }g= dg + Ag - gB \, .
\ee
The action can be usefully expressed as
\bea
S(A,B;g) &=& S(g) + \frac{1}{4\pi}\int_{\S}B(i+*)g^{-1}dg
+ \frac{1}{4\pi}\int_{\S}A(i-*)dg g^{-1} \nonumber \\
& & \; \;  - \frac{1}{8\pi}\int_{\S}\left( B*B + A*A \right)
  + \frac{1}{4\pi}\int_{\S}B(i +*)g^{-1}Ag . \label{lraction}
\eea
In terms of local co-ordinates, this also fixes our conventions, this reads,
\bea
S(A,B;g)&=& S(g) + \frac{1}{2\pi}\int_{\S}d^{2}z \Tr B_{\zb}g^{-1}\partial_{z}
g - \frac{1}{2\pi}\int_{\S}d^{2}z \Tr A_{z}\partial_{\zb}g .g^{-1}
\nonumber \\
& & \; \; + \frac{1}{2\pi}\int_{\S}\Tr A_{z}gB_{\zb}g^{-1} -
\frac{1}{4\pi}\int_{\S}d^{2}z \Tr \left( A_{z}A_{\zb} + B_{z}B_{\zb}
\right)
\eea

As we noted before, the action (\ref{lraction}) is not gauge invariant
in general, but it is the next best thing. Under the transformations,
\be
A \rightarrow A^{h}= h^{-1}Ah + h^{-1}dh  , \; \; \; B \rightarrow
B^{l} = l^{-1}B l
+ l^{-1}dl , \; \; \; g \rightarrow h^{-1}gl
\ee
we find that

\be
S(A,B;g) \rightarrow S(A,B;g) +  \frac{i}{4\pi}\int_{\S}\left(Bdl.
l^{-1}-Adh . h^{-1} \right)  -i \Gamma(l)+i\Gamma(h).
\ee 
The attractive feature here is that the variation of the action depends
neither on the metric nor on the group element $g$.

There are various ways to construct a gauge invariant action, that is to
satisfy (\ref{anomfree}). A simple choice is the standard diagonal
embedding, $H_{L} \times H_{R} = H
\times H$ for $H \subseteq G$, with $(A,B) = (A,A)$.
Somewhat more exotic is
the case where, if $H=H' \times U(1)$, we can have a left connection
$A' + a$ and a right connection $B' + b$ such that
\be
A'=B'\;\;,\;\;\;\;\;\;a=-b \, ,
\ee
which corresponds to the so called axial gauging of the coset model.

\subsection{Holomorphic Factorization of Bosonic Coset Models}

While the action (\ref{lraction}) is not invariant under gauge
transformations, we may,
nevertheless, use it to create a wavefunction which transforms in a well
prescribed way. Let us consider the wavefunction
\be
\Psi(A,B) = \int Dg \, e^{-kS(A,B;g)} \;\;.  \label{wf}
\ee
Under a gauge transformation, $\Psi$ transforms as

\bea
\Psi(A^{h},B^{l})& =& \ex{ik\Phi(A;h)-ik\Phi(B;l)} \Psi(A,B) \nonumber \\
&=& \ex{- \frac{ik}{4\pi}\int_{\S}\left(Bdl.
l^{-1} - A dh . h^{-1}\right)  +i \Gamma(l)-i\Gamma(h) } \Psi(A,B) .
\eea 
The phase factors $\Phi(A;h)$ and $\Phi(B,l)$ are cocycles.
E.g.\ with $h' \in H_{L}$, $\Phi(A,h)$ satisfies

\be
\Phi(A;hh') = \Phi(A^{h};h') + \Phi(A;h) \;\;,
\label{cocy1}
\ee
and likewise for $B$. 

The $\Psi(A,B)$ are therefore correctly thought of as sections of a
bundle, which we now determine.

Under an infinitesimal set of transformations
\be
\d A = d_{A}u  , \; \; \; \d B= d_{B}v  , \; \; \; \d g = - ug +gv ,
\ee
with $u \in \lh_{L}$ and $v \in \lh_{R}$ the variation of the action is
\be
\d S(A,B;g) =  \frac{i}{4\pi}\int_{\S}\Tr \left(udA-vdB \right),
\label{vartriv} 
\ee
so that $\Psi$ satisfies
\bea
\left( D^{A}_{\mu}\frac{\d}{\d A_{\mu}} + \frac{ik}{4\pi}\eps^{\mu \nu}
\partial_{\mu} A_{\nu} \right)
\Psi(A,B)&=&0 \nonumber \\
\left(D^{B}_{\mu}\frac{\d}{\d B_{\mu}} - \frac{ik}{4\pi}\eps^{\mu \nu}
\partial_{\mu}B_{\nu} \right) \Psi(A,B) &=& 0 \;\;, \label{gtransf}
\eea
where the covariant derivatives are
\be
D^{A}_{\mu} = \partial_{\mu} + [A_{\mu} ,.] \; \; \; D^{B}_{\mu} =
\partial_{\mu} + [B_{\mu},.]\;\; .
\ee
We also have the equations
\bea
\left( \frac{\d}{\d A_{\zb}} - \frac{k}{4\pi}A_{z} \right) \Psi(A,B) &
=&  0 \nonumber \\
\left( \frac{\d}{\d B_{z}} - \frac{k}{4\pi}B_{\zb} \right) \Psi(A,B)&=&0 .
\label{hol}
\eea

Introducing the operators
\bea
\frac{D}{DA_{z}} &=& \frac{\d}{\d A_{z}} + \frac{k}{4\pi}A_{\zb}
\nonumber \\
\frac{D}{DA_{\zb}} &=& \frac{\d}{\d A_{\zb}} - \frac{k}{4\pi}A_{z}
\nonumber \\
\frac{D}{DB_{z}} &=& \frac{\d}{\d B_{z}} - \frac{k}{4\pi}B_{\zb}
\nonumber \\
\frac{D}{DB_{\zb}} &=& \frac{\d}{\d B_{\zb}} + \frac{k}{4\pi}B_{z}\;\;,
\eea
we may rewrite (\ref{hol}) in this new notation as
\bea
\frac{D}{DA_{\zb}} \Psi(A,B) &=& 0 \nonumber \\
\frac{D}{DB_{z}} \Psi(A,B) &=& 0 , \label{holo}
\eea
and (\ref{gtransf}) becomes
\bea
\left( D^{A}_{\mu}\frac{D}{D A_{\mu}} + \frac{ik}{4\pi}\eps^{\mu \nu}
F(A)_{\mu \nu} \right)
\Psi(A,B)&=&0 \nonumber \\
\left(D^{B}_{\mu}\frac{D}{D B_{\mu}} - \frac{ik}{4\pi}\eps^{\mu \nu}
F(B)_{\mu \nu} \right) \Psi(A,B) &=& 0 . \label{lift}
\eea

Geometrically the situation can be described as follows \cite{ewhf}.
Let ${\cal A}\times {\cal B}$ be the space of $(A,B)$ connections on $\S$,
equipped with the symplectic two-form
\be
\o \left((a_{1},b_{1}),(a_{2},b_{2}) \right) = \frac{1}{2\pi}\int_{\S} \Tr
a_{1} \wedge a_{2} - \frac{1}{2\pi}\int_{\S} \Tr
b_{1} \wedge b_{2}. \label{symp2f}
\ee
On ${\cal A}\times {\cal B}$ there is a prequantum line bundle ${\cal L} =
{\cal L}_{1}^{\otimes k} \otimes {\cal L}_{2}^{\otimes (-k)}$, with
${\cal L}_{1}$ a line bundle on ${\cal A}$ and ${\cal L}_{2}$ a line bundle
 over ${\cal B}$.
(\ref{holo}) and (\ref{lift}) taken together tell us that $\Psi(A,B)$ is an
(equivariant)
holomorphic section of the line bundle ${\cal L}$ over ${\cal A}\times
{\cal B}$.

The reason for the interest in these wave functions is that for
$B\in\lg$ the
partition function of the $G/H$ coset model is simply its norm,
\be
Z_{G/H}(\S)= |\Psi|^{2}. \label{sqaction}
\ee
In order to establish this we need to introduce the conjugate of the
wavefunction
\be
\overline{\Psi(A,B)} = \int Dh \, e^{-kS(B,A;h)}.
\ee
with $h \in G$.
We can compute the norm, by performing the Gaussian integral over $B$, with
$B \in \lg$ and $h,l \in G$,
\bea
|\Psi|^{2} & =& \int DA DB Dl Dh \; \ex{-kS(A,B;l) -k S(B,A;h)}  \nonumber \\
& =& \int DA Dl Dh \; \ex{-kS(A; lh)} \nonumber \\
& =& \int DA Dg \; \ex{-kS(A;g)} \;\;,\label{norm}
\eea
which is the desired result. In passing from the first to the second line we
have integrated over $B$ and made use of the Polyakov-Wiegman identity and we
have, as well, normalized the volume
$\int Dh=1$ throughout. Another way of arriving at this result, which
bypasses the need to use the Polyakov-Wiegman identity, is to
notice that as the gauge group associated with $B$ is now all of $G$ one
can gauge fix $l=1$. The dependence on $B$ is now essentially pure
Gaussian with covariance $1$.

Actually, one can overlap the wavefunctions even when $B$ takes values
in some subalgebra of $\lg$  and arrive at
a gauge invariant theory. One may wonder what this theory is. It turns
out to be an anomaly free gauged WZW model with action
\be
S(A,B;g) + S(B,A;h) .
\ee
This model will make sense, with $(g,h) \in G \times G'$ and $(A,B) \in
(\lh , \lh ')$ providing $H$ and $H'$ are subgroups of both $G$ and
$G'$. The coset model is then a $(G \times G')/(H_{L}\times H_{R})$
theory where $H_{L} = (H,H')$ and $H_{R}=(H', H)$. When $H'= G'= G$ this
reduces, as we have seen, to a $G/H$ theory.

In this way we have reproduced a standard coset model and this is
in itself not a particularly interesting observation. It gains
interest, however, when one couples to fermions. We will briefly
come back to this below.
\subsection{Introduction of Observables}
There is a class of observables in $G/H$ coset models
that are of prime importance.
Let $R_{i}$ be an irreducible representation of $G$. The observables in
question are
\be
\bigotimes_{i}\Tr_{R_{i}}(g(x_{i})) .
\ee
Let us denote the correlation function of these by
\be
Z_{G/H}(\{ R_{i} \}; \{x_{i} \} ) .
\ee

In order to establish holomorphic factorization for these correlation
functions we will need to consider a more general class of wavefunctions. Let
$D_{R_{i}}(g)$ be the matrix of the irreducible representation $R_{i}$ of
$G$ acting on the finite dimensional representation space $V_{i}$.
Define the wavefunction
\be
\Psi(A,B, \{ R_{i} \}, \{ x_{i} \} ) = \int Dg \ex{-k S(A,B;g) }
\bigotimes_{i}D_{R_{i}}(g(x_{i}) ) . \label{obswf}
\ee
Notice that these insertions are not gauge invariant. But as the action
is also not invariant this situation is not completely problematic. If
one considers pointed gauge transformations (i.e.\ those that do not act
at the points $\{x_{i}\}$) then the wavefunction transforms just as
in the case without operator insertions considered previously. We
also define a dual wavefunction by
\be
\overline{\Psi(A,B, \{ R_{i} \}, \{x_{i}\} )} = \int Dh \ex{-k S(B,A;h) }
\bigotimes_{i}D_{R_{i}}(h(x_{i}) )
\ee
with the observables positioned at the same points and in the same
representations as in (\ref{obswf}).

Let $\Tr_{R_{i}}(hl)(x_{i})$ denote $\Tr
D_{R_{i}}(l(x_{i}))D_{R_{i}}(h(x_{i}))$. The derivation given in
(\ref{norm}) can be followed through line by line to establish (again
$B \in \lg$)
\be
Z_{G/H}(\{R_{i}\}, \{x_{i}\} ) = \int DB DA \, \bigotimes_{i}
\Tr_{R_{i}}| \Psi(A,B, \{ R_{i}
\} , \{ x_{i} \} ) |^{2}  .
\ee

The new wavefunctions do not obey (\ref{lift}). Let $T_{L,R}^{a,b}$ be the
generators of $\lh_{L,R}$. One finds instead of (\ref{lift})
\bea
&&\left( D^{A}_{\mu}\frac{D}{D A_{\mu}} + \frac{ik}{4\pi}\eps^{\mu \nu}
F(A)_{\mu \nu} \right)
\Psi(A,B; \{R_{i}\}, \{x_{i}\} ) \nonumber \\
&=& -\sum_{i}\d (x-x_{i}) T^{a}_{L} D_{R_{i}} (T^{a}_{L})
\Psi(A,B; \{R_{i}\} , \{x_{i}\} )\nonumber \\
&&\left(D^{B}_{\mu}\frac{D}{D B_{\mu}} - \frac{ik}{4\pi}\eps^{\mu \nu}
F(B)_{\mu \nu} \right) \Psi(A,B; \{R_{i} \}, \{x_{i}\} )  \nonumber \\
&=&  \sum_{i}\d (x -x_{i})   \Psi(A,B; \{R_{i} \}, \{x_{i}\} )
T^{b}_{R} D_{R_{i}}(T^{b}_{R}) ,  \label{lift2}
\eea
with the matrix multiplication understood to be on the appropriate
factors.

\subsection{Variation of the Complex Structure }

So far, we have considered the actions for a fixed complex structure
on $\S$ (entering via the Hodge star $*$ in the action). We will now
look at what happens when one varies the complex structure (or conformal
equivalence class of a metric $\rho$ on $\S$). Thus
let ${\cal S}$ be the space of all conformal classes of metrics on $\S$.
The space of holomorphic and gauge invariant sections of ${\cal L}^{\otimes k}$
depends on the given metric $\rho$; we denote that vector space by
$W_{\rho}$.
As $\rho$ varies over ${\cal S}$, $W_{\rho}$ varies as the fibre of a
vector bundle ${\cal W}$ over the space ${\cal S}$ of complex structures on
$\S$. Thus the wavefunctions we have constructed can be thought of as
sections of ${\cal W}$. Set
\be
\d^{(1,0)} = \int_{\S} \d \rho_{\zb \zb } \frac{\d}{\d \rho_{\zb \zb} }
\, , \; \; \;
\d^{(0,1)} = \int_{\S} \d \rho_{zz } \frac{\d}{\d \rho_{zz} } .
\ee

We notice that\footnote{In subsequent formulae one should
regularize the currents that appear. One way to do this is to use a
gauge invariant point splitting procedure \cite{gawedzki}. The net
effect is that one ought to replace $k$ with $\overline{k} = k +
c_{H}$.}, when $B \in \lg$,
\be
\nabla^{(1,0)}\Psi(A,B;\rho) \equiv \left[ \d^{(1,0)} +
\frac{\pi}{2k}\int_{\S}\d \rho_{\zb\zb} \Tr
\frac{D}{DB_{\zb}} \frac{D}{DB_{\zb} }
\right] \Psi(A,B;\rho) =0 . \label{holstruct}
\ee
This defines an anti-holomorphic structure on ${\cal W}$. Similarly a
holomorphic structure is defined by saying that a section is holomorphic if it
is annihilated by
\be
\nabla^{(0,1)} \equiv  \d^{(0,1)} -
\frac{\pi}{2k}\int_{\S}\d \rho_{z z} \Tr \frac{D}{DA_{z}}\frac{D}{DA_{z}}
\ee
A straightforward exercise shows that if $A\in \lg$ then
\be
\nabla^{(0,1)}\Psi(A,B;\rho)=0. \label{antihol}
\ee

The wavefunctions of the $G/G$ models are therefore special in being
holomorphic and anti-holomorphic. This implies that the norm-squared
of the wave function, i.e.\ the
partition function for the $G/G$ theory, is constant as a function
on $\cal S$ and hence metric independent, defining a topological
field theory. It could have been that the partition function is metric
independent without the stronger statement that it is the norm of a
wavefunction that is both holomorphic and anti-holomorphic.

The operators $\nabla^{(1,0)}$ and $\nabla^{(0,1)}$ will figure
prominently in the proof of metric independence of the topological
models. Indeed, as we will see, for the topological $N=2$ coset models
with $\lh \subset \lg$ the wavefunctions indeed satisfy both
(\ref{holstruct}) and (\ref{antihol}) just as they do for the $G/G$ model.
One can therefore take this trait as a {\em definition} of the
topological coset models.

Altogether (\ref{sqaction}) and (\ref{holstruct})
are the statement of
holomorphic factorization of the $G/H$ coset model. As $B \in \lg$, $\Psi$
can be taken to be an anti-holomorphic section of ${\cal W}$, and
consequently, if $e_{i}(B,\rho)$ form a holomorphic and orthogonal
basis of ${\cal W}$ we can expand $\Psi(A,B;\rho)$ as
\be
\Psi(A,B;\rho)= \sum_{i}\overline{e_{i}(B;\rho)} \Psi_{i}(A,\rho)
\ee
where $\Psi_{i}(A,\rho)$ is anti-holomorphic as a `function' on ${\cal S}$. The
$G/H$ partition function can, therefore, be expressed as
\be
Z_{G/H}(\S;\rho)= \int DA \; \sum_{i} |\Psi_{i}(A;\rho)|^{2} .
\ee
This establishes the holomorphic factorization not only for the standard
diagonal embedding but for all the other anomaly free gauged models
as well, {\em including} the axially gauged theories.

In the case of the $G/G$ model, the wavefunction $\Psi(A,B;\rho)$ must
also be an anti-holomorphic section of ${\cal W}$ so that we may write
it as
\be
\sum_{ij}\overline{e_{i}(B, \rho)}e_{j}(A,\rho) d^{ij}
\ee
where the $d^{ij}$ are numbers. The partition function is then
$|d|^{2}$; an eminently respectable topological invariant. Actually one
can do better and establish that $d^{ij} =\d^{ij}$, so that $Z_{G/G}$
is just the dimension of the space of holomorphic sections of $\cal W$, or
the number of conformal blocks of the $G$ WZW model \cite{ewcp}. This
dimension can be calculated by explicit evaluation of the partition
function \cite{btver}.

The wavefunctions which include observables (\ref{obswf}) are
anti-holomorphic as their insertions introduce
no extra metric dependence. Formally, also other gauge invariant
observables will lead to conformal field theories. For example, the
expectation value of a Wilson loop $\Tr_{R_{i}}P \exp{\oint A}$ will
not spoil (\ref{holstruct}). On the other hand, the introduction of
$\Tr_{R_{i}}P \exp{\oint B}$ will yield a wavefunction that does not
satisfy (\ref{holstruct}). For the $G/G$ model this implies that Wilson
loops of $A$, in spite of the fact that they look like eminently respectable
topological observables, do not lead to topological correlation
functions. This can also
be seen rather directly from the proof of metric independence of the
partition function in \cite{ewhf} (or, for the topological \ks\
models, in section 5.1 below), in which the $A$-equations of
motion enter in a crucial way (e.g.\ in section 5.1 to establish
the BRST-exactness of the energy momentum tensor).

Nevertheless, there are topological observables in the $G/G$ model
depending on $A$ (and $g$), namely the images of `horizontal' Wilson
loops under the equivalence \cite{btver} of Chern-Simons theory on
$\S\times S^{1}$ with the $G/G$ model on $\S$ \cite{bdt}.

\subsection{$N=1$ Coset Models}

The general (anomalously)  gauged WZW model $S(A,B;g)$ studied
above has an $N=1$ supersymmetric extension.
To describe the action and the field content, let us
orthogonally decompose
\be
(\lg_{L,R})^{\CC} = (\lh_{L,R})^{\CC} \oplus (\lk_{L,R})^{\CC}\;\;.
\ee
The supersymmetric extension has Weyl fermions $\psi_{-}$ with values in
$(\lk_{L})^{\CC}$ and $\psi_{+}$ with values in $(\lk_{R})^{\CC}$. The
action is
\be
S(A,B;\psi_{+},\psi_{-};g) = S(A,B;g) + \frac{i}{4\pi}\int_{\S}\psi_{-}
D_{z}(A)\psi_{-}  + \frac{i}{4\pi}\int_{\S}\psi_{+}D_{\zb}(B)\psi_{+} .
\label{supact}
\ee
The covariant derivatives are defined by
\bea
D_{z}(A)\psi_{-}&=& \partial_{z}\psi_{-} + [A_{z}, \psi_{-}] \nonumber \\
D_{\zb}(B)\psi_{+}&=& \partial_{\zb}\psi_{+} + [B_{\zb}, \psi_{+}] .
\eea

The action enjoys the supersymmetry,
\bea
\d g &=& i \eps_{+}\psi_{-}g + i \eps_{-}g\psi_{+} \nonumber \\
\d \psi_{-} & =& \eps_{+}\Pi_{L} (D_{\zb}(B)g.g^{-1} +i\psi_{-}\psi_{-})
\nonumber \\
\d \psi_{+} & =& \eps_{-}\Pi_{R} (g^{-1}D_{z}(A)g -i \psi_{+}\psi_{+}) ,
\label{sup}
\eea
and $\Pi_{L,R}$ projects onto the $\lk_{L,R}$ part of the Lie algebra.
These transformations are completely compatible with the gauge symmetry
of the theory. In particular, this means that one can consider the usual
gauging to arrive at the standard $N=1$ action with $(A,B)=(A,A)$. One may
also consider the axially gauged supersymmetric model $(A'+a,B' +b)=
(A'+a,A' -a)$. However, care must be exercised as the chiral
coupling to the fermions may produce an anomaly. One way around this is
to use, in the $U(1)$ sector, a chirally preserving regularization from the
outset.

When one takes $A=B$, (\ref{supact}) gives one a Lagrangian realization
of the $N=1$ super-GKO construction \cite{gko,schnitzer}.
It is useful to adopt a slightly different notation than that
used in the above equations.
In this situation $A$ is a $\lh\equiv\mbox{Lie}H$ valued gauge field for
the (anomaly free) adjoint subgroup $H$ of $G_{L}\times G_{R}$.
$\p_{\pm}$ are then Weyl fermions taking values in the
complexification $\lk^{\CC}$ of $\lk$, the orthogonal complement
to $\lh$ in $\lg\equiv\mbox{Lie} G$,
\be
\lg=\lh\oplus\lk\;\;,\;\;\;\;\;\;\p_{\pm}\in\lk^{\CC}\;\;.
\ee
We will denote by $\pih$ and $\pik$ the orthogonal projectors onto
$\lh$ and $\lk$ respectively.
The $N=1$ (actually $(1,1)$) supersymmetry in this formulation is,
\bea
\d g &=& i\eps_{-}g \p_{+} + i \eps_{+} \p_{-} g\non
\d \p_{+} &=& \eps_{-} \pik(\g D_{z}g -i\p_{+}\p_{+})\non
\d\p_{-} &=& \eps_{+}\pik (D_{\zb}g \g +i\p_{-}\p_{-})\non
\d A &=& 0\;\;.\label{n1susy}
\eea

\subsection{Supersymmetric Wave Functions and Factorization}

In order to establish the factorization of the general $N=1$
supersymmetric WZW
model, we
follow the same procedure as in the bosonic coset model. Let us begin with the
wavefunction
\be
\Psi^{N=1}(A,B) = \int Dg D\psi_{-} \; \exp{-k\left(S(A,B;g)
+\frac{i}{4\pi}\int_{\S}\psi_{-}D_{z}(A)\psi_{-} \right)} .
\ee
The path integral enjoys the supersymmetry
\bea
\d g &=& i \eps_{+}\psi_{-}g  \nonumber \\
\d \psi_{-} & =& \Pi_{L}\eps_{+} (D_{\zb}(B)g.g^{-1} + i
\psi_{-}\psi_{-}) \label{rsup}
\eea

Unless certain conditions are met the wavefunction vanishes due to the
presence of fermionic zero modes. When there are fermionic zero we should
absorb them by introducing operators of the form of, say
\be
\prod_{i=1}^{n} \psi_{-}(x_{i})
\ee
where $n$ is the number of $\psi_{-}$ zero modes.

Gauge invariance is another story. Not only is the bosonic action not gauge
invariant, but under a gauge transformation the Weyl determinant (i.e.\
the determinant of the Weyl fermions) picks up an
anomaly. Consequently, one has
\be
\left(D_{\mu}^{A}\frac{D}{DA_{\mu}} + \frac{i(k+c_{G})}{4\pi}\eps^{\mu \nu}
F_{\mu \nu}(A) \right) \Psi^{N=1}(A,B) =0
% & & \nonumber \\
%\sum_{i} \d(x-x_{i}) [ T^{a}_{L}& , &\Psi^{N=1}(A,B) ] \;\;,
\ee
and hence a corresponding modification of the geometrical interpretation
of the wave function. This equation picks up commutators on the
righthand side (with delta function support) when operator insertions are
included to soak up zero modes.

We define a conjugate wavefunction by ($h \in G$)
\be
\overline{\Psi^{N=1}(A,B) } = \int Dh D\psi_{+} \; \exp{-k\left(S(B,A;h)
+ \frac{i}{4\pi}\int_{\S}\psi_{+}D_{\zb}(A)\psi_{+} \right)},
\ee
which is invariant under
\bea
\d h &=& i \eps_{-}h\psi_{+}  \nonumber \\
\d \psi_{+} & =& \Pi_{R}\eps_{-} (h^{-1}D_{\zb}(B)h-i\psi_{+}\psi_{+})
 \label{lsup}
\eea

Holomorphic factorization is the the statement that, with $B\in \lg$,
\be
Z^{N=1} = |\Psi^{N=1}|^{2}
\ee
When $B \in \lg$ the fermionic coupling to $B$ in (\ref{supact})
vanishes as there are no $\psi_{+}$'s. Now,
\bea
 |\Psi^{N=1}|^{2}
  & = & \int DA DB \; \Psi^{N=1}(A,B) \overline{ \Psi^{N=1}(A,B)} \nonumber \\
 &=& \int DA D\psi \left(\int DB Dl Dh\, \ex{-k S(A,B;l)-kS(B,A;h)} \right)
\ex{-kS(A, \psi)} \nonumber \\
& =& \int DA D\psi Dg \; \ex{-kS(A,g)-kS(A, \psi)} \label{N=1fac}
\eea
where we have used the factorization of the bosonic coset model in passing
from the second to the third line and defined
\be
S(A,\psi)= \frac{i}{4\pi}\int_{\S} \psi_{-}D_{z}(A)\psi_{-}
+\frac{i}{4\pi}\int_{\S} \psi_{+}D_{\zb}(A)\psi_{+} .
\ee
This is the result that we are after.
In order to see that the supersymmetry variations (\ref{sup}) come out right,
we note that the variable $g$ arises as $g = lh$ and thus
has the transformation rules as in
(\ref{sup}). The Gaussian integration over $B$ is saturated by the equation
of motion
\be
B=- \frac{1}{2}[(i*-1)A^{l}-(i*+1)A^{h^{-1}}]
\ee
which means that the part of the $\psi_{-}$ variation (\ref{rsup})
involving the covariant derivative is
\bea
\d \psi_{-} & =& \Pi_{L}\eps_{+} (D_{\zb}(B)l.l^{-1}\nonumber \\
 & =& \Pi_{L}\eps_{+}( D_{\zb}(A^{h^{-1}})l.l^{-1} \nonumber \\
 & =& \Pi_{L}\eps_{+}( D_{\zb}(A)g.g^{-1}
\eea
which agrees with that in (\ref{sup}). Likewise, from (\ref{lsup})
\bea
\d \psi_{+} & =& \Pi_{R}\eps_{-}h^{-1} D_{z}(B)h \nonumber \\
 & =& \Pi_{R}\eps_{-} h^{-1}D_{z}(A^{l})h \nonumber \\
 & =& \Pi_{R}\eps_{-} g^{-1}D_{z}(A)g ,
\eea
again in agreement with (\ref{sup}).
One can now understand the, rather perplexing, supersymmetry of the
coset models as a symmetry on the left for $l$ and on the right for
$h$.

\subsection{More Supersymmetric Models}

Notice that one gets a perfectly respectable field theory in
(\ref{N=1fac}) even if $B$ takes values in a subalgebra of $\lg$ and is
not coupled to fermions. The
resulting theory will be
both supersymmetric and gauge invariant. The bosonic part of the theory
is, as we noted before, a $(G \times G')/(H_{L} \times H_{R})$ model.
Supersymmetry requires that $T_{e}(G'/H) = T_{e}(G/H)$ for the left and
right movers to match. This essentially sets $G=G'$ (up to discrete group
actions). The supersymmetric model is therefore a $(G \times G)/[(H,H')
\times (H',H)]$ theory with the left and right moving fermions taking
values in $\lg/\lh$.

One may consider $B\in\lh'$ also to couple to fermions in a wavefunction
with action (\ref{supact}). The gauge invariant and supersymmetric
theory that one obtains on taking the norm is identical to the one
described in the previous paragraph except that there are, in addition,
left and right moving fermions with values in $\lg/ \lh '$.

%The square of the wave function (on integrating out $B$) gives rise to the
%action
%\bea
%&& S(g) + S(h) + S(A,\psi) \nonumber\\
%&&+ \frac{1}{4\pi}\int_{\S}A[(i-*)dg. g^{-1} + (i+*)h^{-1}dh]
%- \frac{1}{4\pi}\int_{\S}A*A \nonumber \\
%&& + \frac{i}{16\pi} \int_{\S}[\Pi_{\lh '}\left( (i*-1)(g^{-1}dg +
%g^{-1}Ag) - (i*+1)(hAh^{-1}-dh. h^{-1} \right)]^{2} \;\;.
%\eea
%At one extreme, $B \in \lg$ one
%reproduces {\em the} $N=1$ $G/H$ coset model, while at the other extreme,
%$B=0$, one finds instead that the gauge field $A$ may be eliminated
%algebraically.

\section{$N=2$ Cosets and Topological Kazama-Suzuki Models}

Now we have come to the main part of this paper, in which we will
deal with $N=2$ coset models and their topological partners.
For the most part, we
will be interested in the standard $G/H$ supersymmetric models where
$G$ is a compact semi-simple Lie group (which we will also
throughout assume to be simply laced), and $H$ is a closed subgroup
of $G$.

\subsection{Lagrangian Realization of \ks\ and More General $N=2$ Models}

In \cite{ks}, Kazama and Suzuki
investigated (at the current algebra level), under
which conditions on $G$ and $H$
the $N=1$ superconformal algebra of the coset model
could be extended to an $N=2$ superconformal algebra.
For our purposes, the most convenient characterization of the results
is the following (see e.g.\ \cite{ks,schweigert,jfcs}).
The coset model has an $N=2$ superconformal algebra
iff there exists a direct sum decomposition
\be
\lg^{\CC} = \lh^{\CC}\oplus(\lk^{+}\oplus\lk^{-})
\;\;,
\ee
such that
\bea
1) && \dim \lk^{+} = \dim \lk^{-}\non
2) && [\lk^{\pm},\lk^{\pm}]\ss\lk^{\pm}\non
3) && \Tr|_{\lk^{+}}=\Tr|_{\lk^{-}}=0\;\;.\label{kscon}
\eea
In fact, it can be seen rather directly that
these conditions imply \cite{ewcp,nak} that the
supersymmetry (\ref{n1susy}) of the action
is enlarged to a $(2,2)$ supersymmetry. Namely, denoting
the $\lk^{+}$-components of the fermions $(\p_{+},\p_{-})$ as
$(\a_{+},\b_{+})$ and the $\lk^{-}$-components by $(\b_{-}, \a_{-})$ the
action can be written as
\be
kS_{KS}(g,A,\a,\b) = kS_{G/H}(g,A) +
\frac{ik}{2\pi}\int_{\S} \b_{-}D_{\zb}\a_{+} +\b_{+}D_{z}\a_{-}
\;\;.\label{ksact}
\ee
There is thus an R-symmetry of the fermionic part of the
action with respect to which
the fermions $\a_{\pm}$ and $\b_{\pm}$ have charges $\pm 1$
respectively, and the supersymmetry transformations may be
split into their $\lk^{\pm}$-parts which are separate invariances
of the action. For instance, for the left-movers one has \cite{heneg}
\bea
\d g &=& i\bar{\eps}^{-}g\a_{+} + i\bar{\eps}^{+}g\b_{-}\non
\d \a_{+} &=& \bar{\eps}^{+} \pip(\g D_{z}g -i\a_{+}\b_{-} - i\b_{-}\a_{+})
 -i\bar{\eps}^{-} \a_{+}\b_{-}\non
\d \b_{-} &=& \bar{\eps}^{-} \pim(\g D_{z}g -i\a_{+}\b_{-} - i\b_{-}\a_{+})
 -i\bar{\eps}^{-} \b_{-}\a_{+}\non
\d A &=& \d\a_{-}\, = \d\b_{+} \, = 0\;\;.\label{n2susy}\;\;,
\eea
where $\Pi_{\pm}$ denotes the projectors onto $\lk^{\pm}$ respectively.
There is an analogous equation for the right-mnovers.

The conditions to be satisfied so that the $N=1$ supersymmetry is
enhanced to an $N=2$ supersymmetry can also be met when the gauging is
not diagonal. In fact, when (\ref{kscon}) is satisfied the action
(\ref{supact}) has an $N=2$ supersymmetry. Once more this means that
there are $N=2$ axially gauged models as well as a large class of
non-diagonal coset $N=2$ theories that come on `squaring'
(\ref{supact}). As these models possess the full $N=2$ supersymmetry,
they differ from the `heterotic coset models' studied in \cite{cjetal}
as generalizations of \ks\ models. In the following, we will concentrate
on the standard \ks\ models.

The conditions (\ref{kscon}) imply
in particular that $G/H$ is even dimensional
so that $\rk G - \rk H =2n$ is even.
Regarding the decomposition $\lk^{\CC}=\lk^{+}\oplus\lk^{-}$ as a
decomposition of the complexified tangent space of $G/H$ at the
origin, (\ref{kscon}) implies that $G/H$ has an $H$-invariant
complex structure, the (integrable) $\pm i$ eigenspaces of the
complex structure corresponding to $\lk^{\pm}$ respectively. This means
that we can regard $\lk^{-}$ as the complex conjugate of $\lk^{+}$.

While there are plenty of \ksm s with $\rk H < \rk G$ (e.g.\ based on
even-dimensional groups $G$ with $H$ trivial \cite{sstvp,agew}), it
appears that for most conformal field theory purposes one can restrict
oneself to the case $\rk G = \rk H$ by the sequential $G/H$ method of
\cite{ks} (which permits one to write
\be
G/H \sim G/(H\times U(1)^{2n}) \times U(1)^{2n}
\ee
at the level of symmetry algebras, expressing the given $G/H$ $N=2$
coset model as the product of a model with $H$ of maximal rank and the
well understood $N=2$ theory based on $U(1)^{2n}$) combined with
Abelian S-duality for the bosonic part of the action (see \cite{agew}).

If $\rk G = \rk H$, then $H$ contains the maximal torus $T$ of $G$ and
the conditions (\ref{kscon}) are equivalent to the requirement
that $G/H$ be a K\"ahler manifold.  By a theorem of Borel,
$H$ is then the centralizer of some (not necessarily maximal) torus of $G$.
In that case, $\lh^{\CC} \oplus \lk^{+}$ is a parabolic subalgebra
of $\lg^{\CC}$, and in terms of the Cartan decomposition
\be
\lg^{\CC} = \lt^{\CC} \oplus_{\a}\lg_{\a}\oplus_{\a} \lg_{-\a}
\;\;,\;\;\;\;\;\;\pm\a\in\Delta^{\pm}(G)\;\;,
\ee
the subalgebras $\lk^{\pm}$ can be realized as the sum of the root
spaces associated with the roots
\be
\Delta^{\pm}(G/H) = \Delta^{\pm}(G)\setminus\Delta^{\pm}(H)
\ee
of $G$ which are not roots of $H$,
\be
\lk^{\pm} = \oplus_{\a}\lg_{\pm\a}\;\;,\;\;\;\;\;\;\pm\a\in\Delta^{\pm}(G/H)
\;\;.
\ee
If $\lk$ is a symmetric subalgebra of $\lg$, i.e.\ such that
\be
[\lk,\lk]\ss\lh\;\;,
\ee
which implies that the algbras $\lk^{\pm}$ are Abelian,
\be
[\lk^{\pm},\lk^{\pm}]=0\;\;,
\ee
then $G/H$ is automatically K\"ahler and is what is known as a hermitian
symmetric space. These spaces are completely classified and examples are
the complex Grassmannians $SU(m+n)/SU(m)\times SU(n)\times U(1)$
and the `twistor spaces' $SO(2n)/SU(n)\times U(1)$.

It is \ksm s based on hermitian symmetric spaces which are in a sense
the easiest to understand and which have received the most attention in
the literature (see e.g.\ \cite{ks,lvw,flmw,lw,dg1,dg2}). They are
also phenomenologically the most appealing models as they have no extra
$U(1)$-symmetries beyond that dictated by the $N=2$ superconformal algebra
\cite{ks} and thus lead to the minimal $E_{6}\times E_{8}$ gauge group
via the Gepner construction.

Let us briefly return to the general case to introduce some more notation
that we will require later on. We denote by $c_{G}$ and $c_{H}$ the
dual Coxeter numbers of $G$ and $H$, $c_{H}$ being understood as a vector
$(c_{i})$ when $H$ has several simple factors $H_{i}$,
and by $\rho_{G}$ and $\rho_{H}$ the Weyl vectors of $G$ and $H$,
\be
\rho_{G,H}=\trac{1}{2}\sum_{\a\in\Delta^{+}(G,H)}\a\;\;.
\ee
Their difference $\rho_{G/H}=\rho_{G}-\rho_{H}$ has the property that
it is orthogonal to all the simple factors of $H$,
\be
\Tr \rho_{G/H}\a =0\;\;\;\;\;\;\forall\;\a\in\Delta^{+}(H)
\ee
and lies in the direction generated by the $U(1)$-current of the
$N=2$ superconformal algebra. In the hermitian symmetric models it
is also the generator of the single $U(1)$-factor of $H$.

Upon bosonization the $N=2$ \ks\ level $k$
coset model can be described as the (bosonic) coset
\be
[(G\times SO(\dim G/H)_{1})/H]_{k} \;\;,
\ee
with $SO(\dim G/H)$ at level 1 representing the bosonized fermions,
and the embedding of $H$ into $SO(\dim G/H)$ being given by the
isotropy representation of $H$ on the tangent space of $G/H$. Here the
level of $H$ is given by $k+c_{G}-c_{H}$ (for simply laced $G$).
For more information on the Lie algebraic aspects of these models
we refer to \cite{ht}.

As the $N=2$ theories are in particular $N=1$ models one can adopt the
wavefunctions of the previous section to establish holomorphic
factorization in these theories as well. One can consider more `refined'
wavefunctions as well, which have explicit dependence on the fermion
fields. However, these are not needed for our present purposes.

\subsection{The Topological Twist of the \ks\ Model}

It is well known that an $N=2$ superconformal field theory
can be twisted to a topological conformal field theory, i.e.\
a theory with a BRST-exact and traceless energy momentum tensor
with traceless superpartner \cite{ewsig,ey,ehy}. In fact,
consider the standard $N=2$ superconformal algebra
\bea
G^{\pm}(z)G^{\mp}(w) &=& \frac{\trac{2}{3}c}{(z-w)^{3}} \pm
\frac{2J(w)}{(z-w)^{2}}+\frac{2T(w)\pm \del_{w}J(w)}{(z-w)}+ \ldots\non
J(z)G^{\pm}(w) &=& \pm \frac{G^{\pm}(w)}{(z-w)}+\ldots \non
J(z)J(w)&=&\frac{\trac{1}{3}c}{(z-w)^{2}}+\ldots \non
T(z)J(w) &=& \frac{J(w)}{(z-w)^{2}} + \frac{\del_{w}J(w)}{(z-w)} + \ldots\non
T(z)G^{\pm}(w) &=& \frac{\trac{3}{2}G^{\pm}(w)}{(z-w)^{2}}
+\frac{\del_{w}G^{\pm}(w)}{(z-w)} + \ldots \non
T(z)T(w) &=& \frac{\trac{1}{2}c}{(z-w)^{4}} + \frac{2T(w)}{(z-w)^{2}}
+\frac{\del_{w}T(w)}{(z-w)} + \ldots\;\;.\label{n2sca}
\eea
It follows that the twisted energy momentum tensor
\be
T_{top}(z) = T(z) \pm \trac{1}{2}\del_{z}J(z)\label{twist}
\ee
satisfies a Virasoro algebra with central charge $c_{top}=0$, that
(for the $+$-sign in (\ref{twist}))
the conformal weights of $G^{+}$ and $G^{-}$ have shifted from their
original value $\frac{3}{2}$ to $1$ and $2$ respectively, and that
\be
T_{top}(z) = \{Q,G^{-}(z)\}\;\;,
\ee
where
\be
Q = \oint G^{+}(z)dz\;\;,\;\;\;\;\;\;Q^{2}=0\;\;.
\ee
The physical states of this twisted theory, defined as the cohomology
of $Q$, can be represented by the chiral primary fields \cite{lvw} of
the original $N=2$ superconformal field theory.

In fact, as indicated there are two possible twistings, the second
corresponding to changing the sign of $J$, and upon putting
together left and right movers, one can obtain two inequivalent
topological field theories, known as the A and B models respectively
\cite{ewab}. In the case of $N=2$ superconformal (i.e.\ Calabi-Yau)
sigma models, the $N=2$ $U(1)$-current $J$ is the R-current of the
theory and the twists (\ref{twist}) can be mimicked at the level of
the action by adding a term of the form
\be
\frac{1}{2}(\omega_{z}J_{\zb} \pm \omega_{\zb} J_{z})\label{oj}
\ee
to the Lagrangian, where $\omega$ is the spin-connection.
The whole effect of this term is to change the
spins of the fermions form $1/2$ to $0$ or $1$. In particular, the
A-model then coincides with the topological sigma model introduced
in \cite{ewsig}.

In the case of \ksm s, the $N=2$ $U(1)$-current is more complicated.
At the current algebra level, it is given by the sum of the
$\rho_{G/H}$-component of the $H$-current (itself having a bosonic and a
fermionic contribution) and the fermion number operator \cite{ks}.
Hence, in this case it is certainly not correct to try to implement
(\ref{twist}) by adding the term (\ref{oj}) to the Lagrangian
(\ref{ksact}). Nevertheless, by analogy with the
topological sigma model, there is an obvious guess as to what the
action of the A-model should be, namely the action (\ref{ksact})
where one regards the fields $\a_{+}$ and $\a_{-}$ as Grassmann odd
scalars, and $\b_{-}$ and $\b_{+}$ as anti-commuting $(1,0)$ and $(0,1)$
forms respectively. This is tantamount to shifting the gauge field by half
the spin connection.

It has been shown by Nakatsu and Sugawara \cite{ns} that this `twist',
which amounts to adding to the action (\ref{ksact}) a term of the
form (\ref{oj}) with $J$ replaced by the R-current of the \ksm,
indeed reproduces the A-twist (\ref{twist}) of the energy momentum tensor
by the $N=2$ $U(1)$-current $J$ at the conformal field theory level.
We know of no short-cut that would establish this directly
and refer to the analysis \cite{ns} for the details of the argument.

We will, however, show in section 5 that the energy momentum
tensor of the theory defined by this twisted Lagrangian is BRST exact
(modulo the equations of motion of the gauge field). That it differs
from the energy-momentum tensor of the action (\ref{ksact}) by the
derivative of the R-current is obvious by construction since all one
has changed is the spin of the fermionic fields.

The B-model, on the other hand, can be obtained by an analogous
twisting of a suitable axially gauged $N=2$ model.
In general, the B-model is of interest, in particular in relation with
issues like mirror symmetry \cite{ewab,henmir}. However,
as the $(a,c)$
ring of many of the \ks\ models we will consider in the following
and in \cite{bhtgr} is known to be trivial \cite{lvw} (unless one
considers orbifold models), we will not pursue this here.

\subsection{Basic Properties of the Topological \ks\ Model}

It follows from the above that the topological \ksm\ can be
described by the action
\be
S_{TKS}(g,A,\ap,\abm,\bm,\bbp)=S_{G/H}(g,A) +
\frac{1}{2\pi}\int_{\S}\bm D_{\zb}\ap + \bar{\b}^{+}_{\zb} D_{z}\abm\;\;.
\label{tksact}
\ee
Here $\ap$ and $\abm$ are Grassmann odd scalars taking values in
$\lk^{+}$ and $\lk^{-}$ respectively, $\bm$ is a
Grassmann odd $\lk^{-}$-valued $(1,0)$-form and $\bar{\b}^{+}_{\zb}$ a
Grassmann odd
$\lk^{+}$-valued $(0,1)$-form and we have absorbed a factor of $i$
into the definition of $\ap$ and $\abm$ in order to prevent a
proliferation of $i$'s in subsequent equations. Also, for notational
ease we will from now on mostly not indicate explicitly the form and Lie
algebra labels and denote the fields simply by $(\a,\b,\oa,\ob)$.

The action is invariant under two BRST-like transformations $Q$ and $\oQ$
with $\d = Q + \oQ$, the remnants of the original $N=2$ supersymmetry
and the counterparts of the left- and right-moving
BRST tranformations of the A-twist of the
corresponding conformal field theory. These transformations are
\bea
Qg &=& g\a\non
Q\a &=& - \a^{2}\non
Q\b &=& \pim(\g D_{z}g - [\a,\b])\non
Q(\mbox{rest}) &=& 0\non
\oQ g    &=&\oa g                     \non
\oQ\oa    &=& \oa^{2}                    \non
\oQ\ob    &=&  \pip(D_{\zb}g \g + [\oa,\ob])      \non
\oQ(\mbox{rest})    &=&   0 \label{qa}
\eea
It follows from (\ref{qa}) that both $Q$ and $\oQ$ are nilpotent,
\be
Q^{2}=\oQ^{2}=0\;\;.
\ee
On the other hand, $\d$ is only nilpotent
on-shell. From the above one finds that  $\d^{2}=[Q,\oQ]$ acts as
\bea
&&[Q,\oQ]g=  [Q,\oQ]\a=[Q,\oQ]\oa = 0\;\;,\non
&&[Q,\oQ]\b=\pim(\g D_{z}\oa g)\;\;,\;\;\;\;\;\;
[Q,\oQ]\ob=\pip(g D_{\zb}\a \g)\;\;,
\eea
so that $\d^{2}=0$ modulo the $\b$ equations of motion.

One other interesting property of these transformations is, that they can be
decomposed further into the sum of a standard nilpotent
BRST like symmetry $s$ acting only
on the fermionic part of the action and a `topological' symmetry $Q_{T}$,
acting only on $g$ and $\b$. Both of these
are separate invariances of the action. E.g.\ one has $Q=s + Q_{T}$ with
\bea
&& s\a=-\a^{2}\;\;,\;\;\;\;\;\;s\b=-\pim[\a,\b] \non
&& Q_{T}g = g\a\;\;,\;\;\;\;\;\;Q_{T}\b= \pim(\g D_{z}g) \non
&& s(S_{TKS}) = Q_{T}(S_{TKS}) = s^{2} = 0\;\;.
\eea
The topological symmetry $Q_{T}$ only squares to zero equivariantly,
as is familiar from cohomological field theories and $s$ acts trivially
if $G$ and $H$ are such that $G/H$ is hermitian symmetric. In that case,
the terms quadratic in the Grassmann odd fields are absent from the
right-hand side of (\ref{qa}).

For later use we record here the $A$-variation of the action of the
$A$-model,
\be
\d_{A}S_{TKS}= \trac{1}{2\pi}\int_{\S}
(J_{z}-[\a,\b])\d A_{\zb}+(J_{\zb} - [\oa,\ob]) \d A_{z}
\label{aavar}\;\;,
\ee
where
\be
J_{z} = \g D_{z} g\;\;,\;\;\;\;\;\;J_{\zb} = - D_{\zb}g \g\;\;,
\ee
are the $G$-currents and of course only their $\lh$-parts
$\pih(J_{z})\equiv J_{z}^{\lh}$ and $J_{\zb}^{\lh}$
contribute in (\ref{aavar}).

\subsection{Wave Functions and Holomorphic Factorization}

The twisted wavefunction is defined by
\be
\Psi^{TKS}(A,B)= \int Dg \, \ex{ -k
S_{T}(A,B;\overline{\b},\overline{\a};g) \ } ,
\ee
where
\be
S_{T}(A,B;\overline{\b},\overline{\a};g)= S(A,B;g) + \frac{i}{2\pi}
\int_{\S} \overline{\b}d_{A}\overline{\a} .
\ee
The action $S_{T}$ is invariant under the two sets of
transformations
\bea
 \overline{Q}_{T}g &=&  \overline{\a}g  \nonumber \\
 \overline{Q}_{T}\overline{\b} & =& \Pi_{+}\left( \frac{(1+i*)}{2}d_{B}g.
g^{-1} \right)
\eea
and
\bea
\overline{s}\, \overline{\a} &=&  \overline{\a}^{2} \nonumber \\
\overline{s}\overline{\b} &=& \Pi_{+} [ \overline{\a} , \overline{\b} ]
\eea
(all other transformations are zero).

Arguments, familiar from previous sections, tell us that this wavefunction is
anti-holomorphic in the appropriate sense
\be
\nabla^{(1,0)}\Psi^{TKS}(A,B,g)=0 . \label{holtks}
\ee
What is not apparent, at first sight, is that $\Psi^{TKS}$ is also
holomorphic,
\be
\nabla^{(0,1)}\Psi^{TKS}(A,B;g)=0 .\label{aholtks}
\ee
To establish this one notes that
\be
\nabla^{(0,1)}. \ex{ -kS_{T} } = -[\frac{k}{4\pi}\overline{Q}
\int_{\S} \d * \overline{\b}\left( d_{B}g .g^{-1} + \frac{1}{2}
[\overline{\a} , \overline{\b} ] \right) ]\ex{ -kS_{T} } .
\label{vartks}
\ee
A similar calculation will be performed when we discuss the metric
independence and the coupling to topological gravity in more detail.
There we will be more explicit about the derivation.

By the BRST invariance of the theory the expectation value of
$\overline{Q}$ exact terms is zero. Hence the right hand side of
(\ref{vartks}) vanishes in the path integral and we are thus led to the
fact that the wavefunction is anti-holomorphic as well. As the proof
of factorization goes through {\em verbatim} as well (leading, as in
the bosonic case, to the conclusion that the partition function
is metric independent)
this is about the best that one could hope for in a topological
(conformal) field theory.

So far we have considered topological models coming from a diagonal
coset. However, one may establish the topological nature of a more
general class of theories. E.g. when both $G/H$ and $G/H'$ are K\"{a}hler 
one
can form a twisted version of the, now $N=2$ invariant, action
(\ref{supact}), which we denote by $\hat{S}_{TKS}$. The wavefunctions,
$\hat{\Psi}$, obtained on using $\hat{S}_{TKS}$ as the action are indeed
both holomorphic and anti-holomorphic. One arrives at this conclusion by
the same reasoning we used to show that $\Psi^{TKS}$ is holomorphic.

\section{Localization and Abelianization}

In \cite{ewcp}, Witten has shown that the BRST symmetry of the
topological $\cp{1}$ \ksm\ can be used to localize the theory
to a bosonic $U(1)/U(1)$ model (i.e.\ a compact
Abelian BF theory) with some quantum corrections and selection rules
arising from the chiral anomaly of the fermionic sector. Here we will show
that, in general, a $G/H$ model can be localized to a perturbed
$H/H$ model. Heuristically speaking, the BRST symmetry permits one
to linearize the $G/H$-part of the bosonic action, and up to a chiral
anomaly the resulting determinant cancels against that arising form the
integration over the non-zero-modes of the Grassmann odd fields
(taking values in $(\lg/\lh)^{\CC}$).

By the results of \cite{btver}, this theory in turn can be further localized
(Abelianized) to a $T_{H}/T_{H}$ model ($T_{H}$ a maximal torus of $H$),
permitting one to reduce the task of calculating correlation functions in
the non-linear, non-Abelian and supersymmetric topological \ksm\ to a
calculation of correlators in an Abelian bosonic topological field theory.
As we will show in \cite{bhtgr},
this drastic simplification of the problem permits one to be quite explicit
about the structure of the ring of observables in these models and makes
calculations rather straightforward in some cases.

\subsection{Localization: General Considerations}

The idea behind the localization of the path integral is that, taking $Q$ to
be a BRST operator, the Grassmann odd fields can be thought of as
ghosts. Thinking about them in this way leads to a `cancellation'  between
them and would be physical modes. By the supersymmetry (\ref{qa}), one sees
that
the pairing is between certain components of $g$ and $\a$ and
$\overline{\a}$. Essentially one will be able to eliminate the `$G/H$'  parts
of $g$. This is how it works in principle. Unfortunately, in practice
things are somewhat more complicated. Rather than finding a
straightforward cancellation one finds various possible `branches' for the
cancellation to take place. The reason for this plethora of possibilities
rests in the manner in which one projects onto the paired modes.

There are various ways to establish localization in topological theories,
e.g.\ formulated as a BRST fixed point theorem as in \cite{ewcp}. Here
we present an alternative argument which has the virtue of making
the localization quite explicit. Namely, one can add to the
action (\ref{tksact}) a $Q$-exact term enforcing the localization.
Consider, for example,
\bea
Q(\b\g D_{\zb}g) &=& \pim(\g D_{z}g)\pip(\g D_{\zb}g) -
			   \pim([\a,\b])\pip(\g D_{\zb}g)\non
		      &-& \b D_{\zb}(A^{g})\a \;\;,\label{qbg}
\eea
and add this to the action with an arbitrary coefficient $t$,
\be
S_{TKS}\ra S_{TKS}+
\trac{t}{2\pi}Q\int_{\S}\b_{z}\g D_{\zb}g \;\;.\label{saq}
\ee
The partition function and BRST invariant observables do not depend on $t$
(an easy way to see this is that differentiating with respect to $t$ leads
one to evaluate the expectation value of a BRST exact correlation function,
which vanishes by virtue of the BRST invariance of the theory). One is free
to take various limits, $t \ra 0$ giving the original action, while in the
limit $t\ra\infty$ the semiclassical approximation becomes exact. The
additional term preserves the other symmetries
of the theory.

We wish to take the $t \rightarrow \infty$ limit. In this limit the main
contribution to the path integral will come around configurations for which
the additional term (\ref{qbg}) vanishes. The configurations of interest
for us are
\bea
\pim\g D_{z}g &=& 0 \nonumber \\
\pip \g D_{\zb}g &=& 0 . \label{classeq}
\eea
There are various branches of solutions to these equations. One, which we
shall call the main branch, corresponds to $g \in H$ for which
(\ref{classeq}) is automatically satisfied. This branch is algebraic,
meaning that no differential equations are taken to be satisfied. This is
the only branch with this property.
Write
\be
g = \ex{i\f } = \ex{i ( \f^{h} +  \f^{+} + \f^{-})  } .
\ee
The main branch corresponds to a cancellation between $\f^{+}$ and
$\a^{+}$ and between $\f^{-}$ and $\overline{\a}^{-}$. In performing the
path integral, the configuration $g \in H$ will correspond to a
`background' field, while $\f^{\pm}$ are the quantum fields.

There are, however, many other branches. With the help of the Duhamel
formula
\be
\d \ex{X} = \ex{X}\int_{0}^{1}ds \ex{-sX} \d X \ex{sX}\;\;,
\ee
one has
\bea
\pim \left( \frac{\Ad(\ex{i\f})-1}{\ad(\f)} \right) D_{z}\f  &=& 0
\nonumber \\
\pip \left( \frac{\Ad(\ex{i\f})-1}{\ad(\f)} \right) D_{\zb}\f  &=& 0 .
\eea
For example, for the hermitian symmetric spaces, there is a branch with
$\f^{h}=0$ and $D_{z}\f^{-} =D_{\zb}\f^{+}=0$. These equations can only be
satisfied if certain topological criteria are met. This branch will or will
not appear depending on the topology whereas the main branch is always
present.

Even presuming that the bundle is such that these types
of solutions are allowed, as long as we do not consider observables that
include $\a$, this branch cannot contribute, for if there are such $\f$ then
there will be $\a$ zero-modes.

\subsection{Localization onto the Main Branch}

When $t \ra \infty$ the theory localizes onto solutions to $\pim\g
D_{z}g=0$. In principle, for all the branches to be taken into account
we would have to perform a background field expansion around each branch
and in some way handle the fact that the different branches intersect.
See \cite{ewcp} for the corresponding discussion in the case of $\cp{1}$
models, where there is only a rather small number of possible branches,
and where by explicit analysis it can be argued that only the main branch
contributes.

Here, we have no way of (or, at least, we have not succeeded in) eliminating
all the other possible branches in full
generality. Rather, we will proceed with the tacit assumption that only the
main branch, with $g\in H$, contributes to the evaluation of the path
integral and expectation values of the operators that are of interest to us.

When dealing with the localized theory one gets more than just the
classical configurations, of course, as
there are one loop contributions to be
taken into account as well.
To implement this we consider the following scaling of the bosonic fields
$\f^{\pm}$ and
the non-zero-modes of the fermionic fields,
\bea
(\f^{+},\f^{-}) &\ra& (t^{-1/2}\f^{+},t^{-1/2}\f^{-})\non
(\a,\oa) &\ra& (t^{-1/2}\a,t^{-1/2}\oa)\non
(\b,\ob) &\ra& (t^{-1/2}\b,t^{1/2}\ob)\;\;.
\eea
This scaling has unit Jacobian as the transformation of the bosonic fields
is precisely compensated by that of the Grassmann odd scalars.

On using the Duhamel formula one then finds that
\bea
\pim \g D_{z}g &\ra& iD_{z}(A^{h}) \tilde{\f}^{-} + O(t^{-1/2})\;\;,\non
\pip \g D_{\zb}g &\ra& iD_{\zb}(A^{h}) \tilde{\f}^{+} + O(t^{-1/2})\;\;,
\label{jpm}
\eea
where $h= \exp i\f^{h}$ and the twisted fields $\tilde{\f}^{\pm}\in\lk^{\pm}$
are defined by
\be
\tilde{\f}^{\pm} = \int_{0}^{1}ds \ex{-is\f^{h}}\f^{\pm}\ex{is\f^{h}} =
i\left(\frac{\ex{-i \ad \f^{h} }-1 }{\ad \f^{h}} \right)
\f^{\pm}. \label{tilf}
\ee

One question that arises is, what happens to the path integral measure in
passing from the group $G$ valued field $g$ to the group $H$ valued field
$h$ and the $G/H$ coset fields $\tilde{\f}^{\pm}$? On making use of the
Duhamel formula, we find that
\be
Dg = \frak{j}_{\lg}(\f)D\f
\ee
where
\be
\frak{j}_{\lg}(\f) = \Det_{\lg}\left( \frac{1-\ex{-\ad i \f}}{i\ad \f}
\right).
\ee
Now on passing to the $\tilde{\f^{\pm}}$ fields,
\be
Dg = \frak{j}_{\lg}(\f)  [j_{\lg / \lh}(\f^{h})]^{-1} D\f^{h} D\tilde{\f}^{+}
D\tilde{\f}^{-}
\ee
with
\be
\frak{j}_{\lg /\lh}(\f^{h})= \Det_{\lg /\lh} \left(\frac{\ex{-i \ad \f^{h}
- 1} }{\ad
\f^{h}} \right) .
\ee

Thus, the following things happen as $t\ra\infty$:
\begin{itemize}
\item The bosonic action $S_{G/H}(g,A)$ reduces to the action
      $S_{H/H}(h,A)$ of the bosonic topological $H/H$ model.
\item The second term of (\ref{qbg}) vanishes.
\item The term $\b D_{\zb}\a$ of the original action disappears. It is
      replaced by the third term of (\ref{qbg}), which reduces to
      $\b D_{\zb}(A^{h})\a$. The other fermionic term remains unchanged.
\item The first term in (\ref{qbg}) only receives a contribution form the
      order $t^{-1/2}$ terms (\ref{jpm}) of $\pim(\g D_{z}g)$ and
      $\pip(\g D_{\zb}g)$. Thus the kinetic term for $\tilde{\f}^{\pm}$ is
     \[ D_{z}(A^{h})\tilde{\f}^{-}D_{\zb}(A^{h})\tilde{\f}^{+}\;\;.\]
\item The path integral measure becomes $Dg = \frak{j}_{\lh}(\f^{h})D\f^{h}
       D\tilde{\f}^{+} D\tilde{\f}^{-} = Dh D\tilde{\f}^{+}
       D\tilde{\f}^{-}$ (the $t$ dependence of the scaling has already been
       cancelled against the scaling of the Grassmann odd scalars).
\end{itemize}
Putting everything together one finds that the $t\ra\infty$ limit of
(\ref{saq}) is
\be
S_{H/H}(h,A) + \trac{1}{2\pi}\int_{\S}\b D_{\zb}(A^{h})\a +
\ob D_{z}(A)\oa + D_{z}(A^{h})\tilde{\f}_{-}D_{\zb}(A^{h})
\tilde{\f}_{+}\;\;,
\ee
with the canonical measure for all the fields.
Were it not for the fact that it is the operator $D_{z}(A)$ rather than
$D_{z}(A^{h})$ which acts on $\oa$, the determinants from the bosonic and
fermionic terms would cancel modulo questions related to $\b$ zero modes.
This can be made a little bit more explicit by  adding a term $\epsilon
\b\ob$ to the action and integrating out $\b$ and $\ob$. If there
are no $\b$ zero modes, then integrating over the $\a$ first implies $\b
=\ob = 0$, so that everything is independent of $\epsilon$. If there
are $\b$-zero modes, then these have to be soaked up and this can be
accomplished by adding an analogous term $(\b)^{0}(\ob)^{0}$ to the
action (see section 6.4) and the argument still goes through. Anyway,
upon integrating out the $\b$'s, one finds that the $G/H$-part of the
action reduces to
\be
\tilde{\f}^{+}D_{\zb}(A^{h})D_{z}(A^{h})\tilde{\f}^{-}
+ \a D_{\zb}(A^{h})D_{z}(A)\oa\;\;.
\ee
Thus, finally, using the gauge invariance of the chiral anomaly and the
$H/H$ action ($S_{H/H}(h,A) = S_{H/H}(h,A^{h})$) to send $A^{h}$ to $A$,
one finds that the $G/H$-part of the topological \ksm\ (the transverse part
of the partition function) reduces to the ratio of determinants
\be
Z_{\mbox{trans}}= \ex{-W(A,h)} =
\frac{\Det_{\lk} [D_{\zb}(A)D_{z}(A^{h^{-1}})]}%
{\Det_{\lk} [D_{\zb}(A)D_{z}(A)]}\;\;.
\ee
The general structure of the effective action $W(A,h)$, as determined
by the chiral anomalies, is as follows. Let us write the gauge group
$H$ as a product of (semi-) simple factors $H_{i}$ and an Abelian part
$U(1)^{l}$. Then each of the factors $H_{i}$ gives rise to an $H_{i}/H_{i}$
model at level $c_{G}-c_{i}$ where $c_{i}$ is the dual Coxeter number of
$H_{i}$. Combining this with the level $k$ $H_{i}/H_{i}$ actions which
are the remnants of the bosonic $G/H$ action after localization, one thus
obtains, for each factor $H_{i}$, an $H_{i}/H_{i}$ action at level
$k+c_{G}-c_{i}$. This is in agreement with the current algebra (coset
model) description of the \ksm s as (cf.\ section 3) the coset
\be
(G_{k} \times SO_{1}(\dim G/H))/H_{k+c_{G}-c_{H}}\;\;.
\ee
Likewise the gauge field coupling of the $U(1)$-factors
is given by a level $c_{G}$ ($c_{U(1)}=0$) $U(1)/U(1)$ model (i.e.\
essentially a compact Abelian BF theory). However, because in the
twisted model we are dealing
not with fermions but with a $(1,0)$ system, there is an additional
coupling to the scalar curvature $R$ of the spin-connection implicit in
(\ref{tksact}). This coupling takes the form ($h=\exp i\f^{h}$)
\be
S_{TKS}^{\mbox{grav}} = -\trac{i}{2\pi}\int_{\S}\Tr \rho_{G/H}\f^{h} R\;\;,
\label{sgrav}
\ee
where $R$ is normalized such that
\be
\trac{1}{2\pi}\int_{\S}R = \c(\S)
\ee
is the Euler number of $\S$ ($(2-2g)$ for a genus $g$ surface).
As $\rho_{G/H}$ is orthogonal to $\times_{i}H_{i}$, it is indeed
only the $U(1)$-factors of $H$ which contribute to this expression.

Altogether one finds that the effective action
one obtains after localization and integration over the coset valued
fermionic and bosonic fields is
\be
S_{TKS}^{eff,k} = \sum_{i} (k+c_{G}-c_{i})S_{H_{i}/H_{i}}
   + (k+c_{G}) S_{U(1)^{l}/U(1)^{l}} + S_{TKS}^{\mbox{grav}}\;\;.
\ee
It should be borne in mind that, in order to arrive at this action,
we have integrated out only the non-zero-modes of the fermionic
fields, so that the integration over the zero modes is still to
be performed. This is best understood within the context of
specific examples, and we will describe this for the
hermitian symmetric models in \cite{bhtgr}.

Localization can also be performed at the level of wave functions and it
is readily checked that the
result of localizing onto the main branch, at the level of the
wave functions, reproduces the effective action $S_{TKS}^{eff,k}$.

\subsection{Abelianization}

If the gauge group $H$ is Abelian (e.g.\ when one is dealing with the
flag manifolds $G/T$, $T$ a maximal torus of $G$), then the above
localization establishes directly that calculations of correlation
functions can be reduced to calculations in an Abelian topological
field theory, thus significantly simplifying that task.

Something analogous can be achieved in the general non-Abelian case as
well. Indeed,
in \cite{btver} it was shown that the $H_{i}$ gauge symmetry of
the $H_{i}/H_{i}$ model can be used to abelianize the
theory, i.e.\ to reduce it to a $T_{i}/T_{i}$ model (plus quantum corrections)
where $T_{i}$ is the maximal torus of $H_{i}$.
Combining this with the $S_{U(1)^{l}/U(1)^{l}}$ action, one thus obtains
a $T_{H}/T_{H}$ model, where $T_{H}$ is the maximal torus of $H$.
In particular, therefore,
if we are dealing with the K\"ahler models with $\rk G =\rk H$,
then the \ksm\ can be reduced to an Abelian $T/T$ model (with the above
mentioned quantum corrections and the gravitational coupling (\ref{sgrav})).

These quantum corrections are
of two kinds. The first one is a shift of the level of the $H_{i}/H_{i}$
action by $c_{i}$ so that
\be
S_{TKS}^{eff,k} \ra (k+c_{G})S_{T_{H}/T_{H}} + S_{TKS}^{\mbox{grav}}\;\;.
\label{ttac}
\ee

The second correction is a finite-dimensional determinant arising from
the ratio of the functional determinant from the $H_{i}/T_{i}$ components of
the gauge field and the Jacobian (Faddeev-Popov determinant) from the
change of variables (choice of gauge) $h_{i}\ra t_{i}\in T_{i}$.
It can be written as
\be
\exp\left[\trac{1}{4\pi}\int_{\S}R \log\det(1-\Ad(t_{i}))\right]
\;\;,
\ee
where the determinant is taken on
the orthogonal complement of the Lie algebra of $T_{i}$
in the Lie algebra of $H_{i}$. While we refer to \cite{btver} for a detailed
derivation, let us make the following comments here.

\begin{enumerate}
\item
The determinant is what is known as the Weyl determinant
$\Delta_{W}^{H_{i}}(t_{i})$ appearing in the finite-dimensional
Weyl integral formula relating integrals of class functions on $H_{i}$ to
integrals over $T_{i}$.
\item
As we are dealing with a topological theory, we can treat the fields
$t_{i}$ as position independent (in fact, explicit calculations show
that only the constant modes of $t_{i}$ contribute to the path integral).
\item
Hence we can say that the localized and abelianized theory is defined by
the action (\ref{ttac}) with the measure for $t\in T_{H}$ modified to
\be
D[t]\ra D[t](\Delta^{H}_{W}(t))^{\c(\S_{g})/2}\;\;.\label{weylm}
\ee
\end{enumerate}

We have thus managed to reduce the original non-Abelian supersymmetric
topological \ksm\ to a much more tractable bosonic Abelian topological field
theory, the entire information on the coset valued fields being encoded
in the shifted levels and the gravitational coupling, and the Weyl determinant
keeping track of the originally non-Abelian nature of the theory.

In performing actual calculations in the model described above,
there are some further things that are good to keep in mind, for instance
that use of the infinite-dimensional version of the Weyl integral formula
employed above engenders \cite{btdia} a sum over all isomorphism classes
of $T_{i}$-bundles on $\S$ \cite{btver}. This as well as questions related
to chiral anomalies, fermionic zero modes and the ensuing selection rules
for correlation functions, which are more or less immediate analogues of
those considered by Witten \cite{ewcp} for the $\cp{1}$ model, will be
explained for the hermitian symmetric models in \cite{bhtgr}.

\section{Coupling to Topological Gravity}

In this section we will construct the coupling to topological gravity
of the topological \ksm s introduced above. As a preliminary first
step we prove by direct calculation that the partition function of
the pure matter theory is indeed metric independent and that the
energy-momentum tensor is BRST exact modulo the gauge field equations
of motion (i.e.\ modulo the gauge currents). With this in hand one can, to
some extent, rationalize the form of the theory coupled to topological
gravity.

\subsection{Metric Independence of the Partition Function}

The topological nature of the theory defined by the
action (\ref{tksact}) follows (indirectly) from the fact that
at the conformal field theory level this theory is equivalent
to the A-twist of the \ksm. It can also be inferred from the fact that the
wave functions are both holomorphic and anti-holomorphic. However, it is
instructive to establish this
metric independence directly at the level of the Lagrangian realization
we have chosen (this derivation also provides some of the steps required in
showing that the wave functions are anti-holomorphic).

As usual, the energy-momentum tensor is defined as the variation of the
action with respect to the metric $\rho_{\mu\nu}$, i.e.
\be
\d_{\rho}S = \int \sqrt{\rho} \d \rho^{\m\n}T_{\m\n}\;\;.
\ee
Applied to the action (\ref{tksact}), one finds that (as expected)
$T_{\mu\nu}$ is traceless, $T_{z\zb}=0$ and that the non-vanishing
components are
\bea
2\pi T_{zz} &=& -\trac{1}{2}\Tr J_{z}J_{z} + \Tr \b_{z}^{-}D_{z}\a^{+}\non
&=& -\trac{1}{2}\Tr J_{z}^{\lh} J_{z}^{\lh}
    -\Tr J_{z}^{+} J_{z}^{-} + \Tr \b_{z}^{-}D_{z}\a^{+}
\label{tzz}\\
2\pi T_{\zb\zb} &=&
 -\trac{1}{2}\Tr J_{\zb}^{\lh} J_{\zb}^{\lh}
    -\Tr J_{\zb}^{+} J_{\zb}^{-} + \Tr \bb_{\zb}^{+}D_{\zb}\ab^{-}
\eea
Here $J_{z}$ and $J_{\zb}$ are the bosonic currents,
$J_{z}=\g D_{z}g$, $J_{\zb}=- D_{\zb}g \g$, and $J^{\lh}$ and $J^{\pm}$
denote their components in $\lh$ and $\lk^{\pm}$ respectively.

The bosonic part of the energy momentum tensor
is just of the (covariantized) Sugawara
form, while the fermionic part is the standard energy-momentum tensor of
a $(1,0)$-system.

We will now establish that, modulo the $\azb$-equation of motion,
the left-moving energy momentum tensor
$T_{zz}$ is $Q$-exact. Completely analogously, one can establish that
$T_{\zb\zb}$ is $\oQ$-exact, where we have used the decomposition
$\d = Q + \oQ$ of the BRST symmetry into its left- and right-moving
parts. By standard arguments, this then establishes the metric independence
of the partition function and suitable correlation functions (e.g.\
correlation functions of BRST invariant and metric and $A$-independent
operators).

Let us recall the action of $Q=Q_{T}+s$ on the fields,
\bea
Qg &=& g\a\nonumber\\
Q\a^{+} &=& -\trac{1}{2}[\a^{+},\a^{+}]\non
Q\b_{z}^{-} &=& J_{z}^{-} -\pim [\a^{+},\b_{z}^{-}]\non
Q(\mbox{rest}) &=& Q^{2} =0\;\;.
\eea
(here $s$ gives rise to the terms quadratic in the fermions).
and the $\azb$ equation of motion expressing the vanishing of the $H$
gauge current,
\be
J_{z}^{\lh} = \pih [\a^{+},\b_{z}^{-}]\;\;.\label{hcur}
\ee
One can use this equation to replace the term in (\ref{tzz})
quadratic in the bosonic part of the gauge current by a quartic
fermionic term,
\be
J_{z}^{\lh}J_{z}^{\lh}\ra
\pih [\a^{+},\b_{z}^{-}]\pih [\a^{+},\b_{z}^{-}]\;\;.    \label{jhjh}
\ee
This term, which vanishes identically for the hermitian symmetric models, is
in general $s$-exact,
\be
s(\Tr \b_{z}^{-}[\a^{+},\b_{z}^{-}]) = \Tr
\pih [\a^{+},\b_{z}^{-}]\pih [\a^{+},\b_{z}^{-}]\;\;.
\ee
The fact that this part of the metric variation can be cancelled by an
expression quadratic in the gauge currents we have encountered before
in the guise of the operator $\nabla^{(1,0)}$ (cf.\ (\ref{holstruct})).

Furthermore, one can replace $J_{z}^{-}$ by $Q_{T}\b_{z}^{-}$.
Proceeding in this way, one finds that
\be
2\pi T_{zz} = - Q\Tr \b_{z}^{-} (J_{z}^{+}-\trac{1}{2}[\a^{+},\b_{z}^{-}])\;\;.
\label{tqbj}
\ee
To establish this directly, one calculates (a trace will be understood
in the following)
\bea
Q\b_{z}^{-}J_{z} &=& J_{z}^{-}J_{z}^{+}-[\a^{+},\b_{z}^{-}]^{-}J_{z}^{+}
		 + [\a^{+},\b_{z}^{-}]J_{z} -\b_{z}^{-}D_{z}\a^{+}\non
&=& J_{z}^{-}J_{z}^{+} -\b_{z}^{-}D_{z}\a^{+}\non
		 && + \pih [\a^{+},\b_{z}^{-}]\pih [\a^{+},\b_{z}^{-}]
		    + \pip [\a^{+},\b_{z}^{-}]J_{z}^{-}
\eea
(where one has used (\ref{hcur}) in the second step), and
\be
-\trac{1}{2}Q\b_{z}^{-}[\a^{+},\b_{z}^{-}] =
-J_{z}^{-}\pip [\a^{+},\b_{z}^{-}] -\trac{1}{2}\pih [\a^{+},\b_{z}^{-}]
\pih [\a^{+},\b_{z}^{-}]\;\;.
\ee
Thus, upon putting the two together and using (\ref{jhjh})
one finds (\ref{tqbj}). The same argument
establishes (\ref{vartks}) and hence the holomorphicity
(\ref{aholtks}) of the twisted wave function $\Psi^{TKS}$
we studied in section 3.4. In fact, using the holomorphic factorization
of the partition function and `pulling apart' the above argument, taking
care to turn ordinary $A$-derivatives into covariant derivatives once they
start acting on wave functions, one recovers the formulae
(\ref{holtks},\ref{aholtks}).

\subsection{The Coupling to Topological Gravity}

Witten considered the coupling of the twisted $SU(2)/U(1)$ model to
topological gravity in \cite{ewcp}. He found the coupled model by using
the ``Noether procedure''. We will present the coupling of an arbitrary
twisted Kazama-Suzuki model to topological gravity. The informal proof,
above, of the metric independence of the partition function will allow us
to gain extra insight to the final form of the action even though we also have
arrived at the action by a step by step process.

The models that we present here represent a large new class of topological
matter theories coupled to topological gravity (and hence of topological
string theories).

The topological gravity multiplet is taken to be composed of the metric
$\rho_{\mu \nu}$ and its superpartner $\chi_{\mu \nu}$, which transform as
\bea
\d \rho_{\mu \nu}&=&\chi_{\mu \nu} \nonumber \\
\d \chi_{\mu \nu}&=&0 . \label{gravmult}
\eea
Conventionally one needs to add a commuting vector $C^{\mu}$, representing
the reparameterization invariance. However, as argued by Witten, at the
classical level such ghosts are zero and that is all that matters.

To make
the writing of the action as simple as possible, we take the metric
variation to be $\d *$ which is to be understood as a variation of the
complex structure $J^{\nu}_{\;\; \mu}$. This
means, for example, that for a one form $*\o = \o_{\nu}J^{\nu}_{\;\;
\mu}dx^{\mu}$, the metric variation is
\be
\d (*\o) \equiv (\d*)\o \equiv \o_{\nu} \d J^{\nu}_{\;\; \mu} dx^{\mu} .
\ee
For a self dual one form, $\b$, satisfying $(1+i*)\b=0$, there is a hidden
metric dependence so that one has
\be
\d \b = -\frac{i}{2}(\d*)\b
\ee
which is perpendicular to $\b$, that is, $(1-i*)\d \b =0$.

With our conventions the relationship between the variations
(\ref{gravmult}) and $\d*$ are
\bea
\d J^{z}_{\;\; \overline{z}}& =& - \rho^{z\overline{z}} \chi_{\overline{z}
\overline{z}} \nonumber \\
\d J^{\overline{z}}_{\;\; z} &=& \rho^{\overline{z}z}\chi_{zz} .
\eea

The action which includes the coupling to topological gravity is
\bea
&&S_{TKS+TG} = S_{TKS} + \frac{1}{4\pi}\int_{\S} \b \d* g^{-1}d_{A}g +
\frac{1}{4\pi}\int_{\S} \overline{\b}\d* d_{A}g . g^{-1}\non
&&+
\frac{i}{8\pi}\int_{\S}\d* \overline{\b} \d* g \b g^{-1}
+ \frac{i}{8\pi}\int_{\S} [ \overline{\b} , \overline{\a} ] \d *
\overline{\b} - \frac{i}{8\pi}\int_{\S} [\b , \a ] \d * \b .
\eea
Apart from the last two terms this action is the same as (or, rather,
the obvious generalization of) that for the
$\CC\PP^{1}$ model. The reason for the additional terms stems from the fact
that for a generic coset model the anticommutator of the ghost fields in
the ghost transformation rules (\ref{qa})
is not zero. For the hermitian symmetric
cosets the commutators vanish and the action simplifies
accordingly.

The transformation laws which leave the action invariant are
\bea
\d g &=&  \eps g \a +  \eps \overline{\a}g \nonumber \\
\d \a &=& -  \eps \a^{2} \nonumber \\
\d \overline{\a} &=&  \eps \overline{\a}^{2} \nonumber \\
\d \b &=& \eps \Pi_{-}\frac{(1-i*)}{2}\left( g^{-1}d_{A}g -[ \a , \b ] -
\frac{i}{2} \d*
g^{-1} \overline{\b}g \right) \nonumber \\
\d \overline{\b} &=& \eps \Pi_{+}\frac{(1+i*)}{2}\left(d_{A}g. g^{-1} +[
\overline{\a} ,
\overline{\b} ] + \frac{i}{2} \d*
g\b g^{-1}  \right) \nonumber \\
\d A &=& \frac{\eps}{4}\Pi_{\lh}\left ( \frac{(1-i*)}{2}i\d* (d_{A}g.g^{-1}
+ [\overline{ \a} , \overline{\b} ] ) +
\frac{(1+i*)}{2}i\d * (g^{-1}d_{A}g - [\a, \b ] )
\right. \nonumber \\
 & & \; \; \; \; \left. -\frac{1}{2}\d* \d*
(g\b g^{-1}-g^{-1}\overline{\b}g)  \right)
\eea

These are somewhat unedifying, though one should notice that for the hermitian
symmetric cosets these transformations coincide with those found for
the $\cp{1}$ model in \cite{ewcp}.

In order to facilitate the checking of the
invariance of the action it is best to express $S_{TKS}$ in the form
\be
S_{TKS}= S_{G/H} - \frac{i}{2\pi}\int_{\S}\frac{(1-i*)}{2} \b d_{A}\a
+\frac{i}{2\pi}\int_{\S}\frac{(1+i*)}{2} \overline{\b} d_{A}\overline{\a} .
\ee

Let us compare the action and transformation rules with the considerations
of the previous section. The transformation rules for $g$, $\a$ and
$\overline{\a}$ are as before and require no comment. The extra term in the
$\b_{z}$ variation is there, as we noted before, to ensure that it remains
self dual with respect to the deformed metric. The $A_{\zb}$
equation of motion was required to eliminate the $J_{z}^{\lh}J_{z}^{\lh}$
part of the stress tensor (\ref{tzz}). The variation of the action gives,
up to fermionic terms, $\d A_{\zb} J_{z}^{\lh}$, so the appearance of
$J_{z}^{\lh}$ in the transformation rule $A_{\zb}$ is explained.

\subsection{Preliminary Remarks on Selection Rules}
At this level of generality it is difficult to make precise statements
about the correspondence between this field theoretic construct and more
algebro-geometric considerations. However, there are some simple
observations that one can make.

\begin{enumerate}
\item
Witten's correspondence
between the field theory and algebraic geometry for the $SU(2)/U(1)$
model can be straightforwardly extended to the hermitian symmetric $G/H$
models.
\item
The fact that the $\f -A$
system is not a natural object in cohomological theories
gave rise to some complications in the analysis of \cite{ewcp}. This
system does not arise in the case of twisted $N=2$ $G$-models,
i.e.\ when $H = \{ e \}$.
\item
The generalization of the superselection rule (3.39) of
\cite{ewcp} is easily obtained.  The right hand side is obtained on
scaling the fields $(\chi^{zz}, \a)  \ra
e^{\la}(\chi^{zz} , \a)$ and $\b \ra e^{-\la} \b $. This transformation,
plus a similar one for the barred fields, is an invariance of the
action. The right
hand side is a measure of how anomalous the transformation is.
\item
The superselection rule (3.41) of \cite{ewcp} depends rather strongly on
the choice of $G$ and of $H$. On the basis of the results of \cite{bhtgr},
it can however readily be worked out for the complex Grassmannian models.
\item E.g.\ for the $\cp{n}$ models one has (see section 6.2) $n$ bosonic
operators $O_{i}$. Denoting their $p$'th gravitational descendant by
$\tau_{p}(O_{i})$, the generalization of the combined selection rule
\cite[(3.42)]{ewcp} for a correlator
\be
\langle\prod_{a=1}^{s} \tau_{p_{a}}(O_{i}(x_{a})^{r_{i,a}})\rangle_{TKS+TG}
\ee
in genus $g$ turns out to be
\be
\sum_{a=1}^{s}p_{a} + \sum_{i=1}^{n}i \sum_{a=1}^{s}\frac{r_{i,a}}{k+n+1}
= \frac{2k +3(n+1)}{k+n+1}n(g-1) + ns\;\;.
\ee
\end{enumerate}

\subsection{Wavefunctions (Again) and Holomorphic Factorization}

We now prove that the topological models coupled to topological gravity can
also
be expressed as the norm of a wavefunction. This might be surprising at first,
as the cross term
\be
\int_{\S}\d * \overline{\b} \d * g \b g^{-1}
\ee
would appear to spoil a direct factorization. Nevertheless, using the freedom
afforded by the extra field $B$, one can show that the partition function is
the norm of a wavefunction.

Indeed let
\be
\Psi(A,B;*,\d *) = \int Dg D\overline{\a} D\overline{\b}\, \ex{-k S(A,B; g;
\overline{\a},\overline{\b}; *,
\d *)}
\ee
where
\be
S(A,B; g;\overline{\a},\overline{\b}; *, \d *) = S(A,B; g;
\overline{\a},\overline{\b}) - \frac{1}{4\pi}\int_{\S} \left( d_{B}g.g^{-1}
+ \frac{1}{2}[ \overline{\b}, \overline{\a}] \right) \d * \overline{\b} .
\ee
The dual wave function is defined by
\be
\overline{\Psi(A,B;*,\d *)} = \int Dg D\a D\b \, \ex{-k
 S(B,A;g; \a,\b ; *, \d *)}
\ee
where
\be
S(B,A; g;\a,\b; *, \d *) = S(B,A; g;
\a,\b) + \frac{1}{4\pi}\int_{\S} \left( g^{-1} d_{B}g
+ \frac{1}{2}[ \b, \a ] \right) \d * \b .
\ee

One now checks that
\be
Z_{TKS}^{\;\;\;TG} = \int DB\, DA |\Psi(A,B;*,\d *)|^{2} .
\ee

\section{Observables}

In the previous sections we have discussed some of the important features
of the functional integral approach to the topological \KS\ models. These
observations will be used in subsequent publications
to evaluate directly and explicitly correlation functions in some of these
models. To that end, we will now describe the observables (physical
operators).

\subsection{Preliminary Remarks}

Observables are local functionals of the fields
(or perhaps integrals thereof)
which are invariant under the symmetries of the theory. In the
present context of topological $G/H$ \KS\ models
this amounts to $H$-gauge invariance and invariance under the BRST-like
supersymmetry $\d$. In fact, the $\d$-invariance of the action implies
by standard arguments that $\d$-exact operators decouple so that
one is actually interested in $\d$-cohomology classes of $H$-invariant
operators.

To some extent, the structure of the observables in these models could
be deduced from the literature on $N=2$ superconformal field theories,
e.g.\ \cite{lvw,ht}. In these works, the chiral ring of \KS\ models
(which becomes the ring of observables of the twisted model)
has been described in terms of the Lie algebra cohomology of affine
algebras. One of the simplifying features of the present action-based
functional integral approach is the possibility to characterize the
observables directly in terms of the finite-dimensional Lie group $G$
and its Lie algebra, loop groups or their Lie algebras never appearing
explicitly.\footnote{This is analogous to the path integral derivation
of the Verlinde formula, deeply rooted in the representation theory of
loop groups, from the $G/G$ model \cite{btver} using only some group
theory of finite-dimensional groups.}

In the following, we will distinguish two classes of operators: purely
bosonic operators and those depending also on the Grassmann odd fields.
The latter are required for non-zero correlation functions whenever
there are fermionic zero modes. Operators soaking up $\beta$
and $\bb$ zero modes can for instance
be constructed from the bosonic operators
using an analogue of
the standard descent-procedure of topological field theories.

Operators involving $\a$'s and $\oa$'s, on the other hand,
have a rather different flavour to them (among other things because
their zero modes can be interpreted rather directly as tangents to the
(left or right) coset space $G/H$). We will only provide a brief
description of these operators here  and discuss them in
detail in \cite{bhtgr}.

Turning therefore to bosonic observables, we will look for
BRST and gauge invariant functionals of the
group valued field $g$. We are deliberately ignoring a possible dependence
of these operators on the connection $A$ as we have seen before that
one needs to use the $A$
equations of motion (Schwinger-Dyson equations) to establish the metric
independence of correlation functions.
We are thus interested in conjugation invariant
functionals $O(g)$ of $g$ invariant under the supersymmetry
\be
\d g = g \a + \oa g\;\;.\label{obs1}
\ee
A rough argument suggests that in all the KS models with
$\rk G=\rk H=r$ one will find $r$ linearly independent bosonic
operators, as there will be $\dim G - \dim H$ conditions imposed by
$\d$-invariance and a $\dim G - (\dim H-r)$ dimensional
space of $H$-invariants.
Below, we will determine these functionals for the hermitian-symmetric models
based on complex Grassmannian manifolds, and then we will discuss
the other prototypical example of full flag manifolds $G/T$ for $G$ a
classical group. Other models of interest could be dealt with along the
same lines.
\subsection{Bosonic Observables for Grassmannian Models}

We will now look for observables in the \KS\ model based on the complex
Grassmannian $G(m,m+n)$ of complex $m$-planes in $\CC^{m+n}$. This
space can be described as the coset
\bea
G/H &=& U(m+n)/U(m)\times U(n) = SU(m+n)/S(U(m)\times U(n))\non
    &\approx& SU(m+n)/SU(m)\times SU(n) \times U(1)\;\;.\label{obs2}
\eea
More explicitly, one embeds $H$ into $G=SU(m+n)$ in block-diagonal form,
\be
h = \mbox{diag}(h^{(m)},h^{(n)})\;\;,\label{obs3}
\ee
where $h^{(m)}$ and $h^{(n)}$ are $U(m)$ and $U(n)$ matrices respectively,
satisfying
\be
\det h^{(m)} \det h^{(n)} = 1\;\;.\label{obs4}
\ee
This amounts to the specification that the single $U(1)$-factor of $H$,
which plays a special role  in the hermitian-symmetric \KS\ models, is
generated by the element $\mbox{diag}(nI_{m},-mI_{n})$ (proportional to
the Weyl vector $\rho_{G/H}$ of $G/H$).

Consequently, the Grassmann-odd scalars $\a$ and $\oa$ have components
\be
\a = (\a_{ij})\;\;,\;\;\;\;\;\;\oa = (\oa_{ji})\;\;,\;\;\;\;\;\;
i=1,\ldots,m\;\;,\;\;j=m+1,\ldots,m+n\;\;.\label{obs5}
\ee
It follows that (with the notation that indices $i_{k}$, $j_{k}$ have the
same range as the indices $i$, $j$ above)
\be
\d g_{i_{1}i_{2}} = 0 \;\;,\label{obs6}
\ee
i.e.\ that the upper left-hand $U(m)$ block $g^{(m)}$ of $g$ is
$\d$-invariant,
\be
\d g^{(m)} = 0 \;\;.\label{obs7}
\ee
The symmetry between $m$ and $n$ is reflected in the fact that there is
also a $U(n)$ block of invariants. Namely. it follows from
\be
\d g^{-1} = - g^{-1} \d g g^{-1} = - (\a g^{-1} + g^{-1}\oa)\;\;,
\label{obs8}
\ee
that
\be
\d (g^{-1})_{j_{1}j_{2}} = 0\;\;,\label{obs9}
\ee
which we will also write as
\be
\d (g^{-1})^{(n)} = 0\;\;.\label{obs10}
\ee
It remains to impose gauge invariance. $H$ gauge transformations
act on $g^{(m)}$ and $(g^{-1})^{(n)}$ by conjugation with $h^{(m)}$
and $h^{(n)}$ respectively. Hence, a complete set of gauge and BRST
invariant operators can be obtained as traces of $g^{(m)}$ and
$(g^{-1})^{(n)}$. One possibility is to consider the traces
\be
\Tr (g^{(m)})^{l}\;\;,\;\;\;\;\;\;l=1,\ldots.m\label{obs11}
\ee
(and likewise for $(g^{-1})^{(n)}$). However, for the cohomological
interpretation of the operators it turns out to be more convenient
to consider as the basic set of operators the traces of $g^{(m)}$ and
$(g^{-1})^{(n)}$ in the exterior powers of the fundamental
representations of $U(m)$ and $U(n)$ respectively. We thus define
\bea
&&O_{l}(g) := \Tr_{\wedge^{l}}g^{(m)}\;\;,\;\;\;\;\;\;l=1,\ldots,m\;\;,
\label{obs12}\\
&&\bar{O}_{l}(g):=\Tr_{\wedge^{l}} (g^{-1})^{(n)}\;\;,\;\;\;\;\;\;
l = 1,\ldots,n\;\;.\label{obs13}
\eea
Since $\det g = 1$, there is one relation between these operators, namely
\be
\det g^{(m)} \equiv O_{m} = \det (g^{-1})^{(n)}\equiv \bar{O}_{n}\;\;.
\label{obs14}
\ee
Altogether, one therefore has $\rk SU(m+n) = m+n-1$ independent
basic gauge and BRST invariant operators generating the ring of observables
of the topological \KS\ model, i.e.\ the chiral-chiral primary ring
of the $G(m,m+n)$ \KS\ model. The ring structure (which will also depend in
a subtle way on the level $k$) can be determined from the
correlation functions of these operators. This will be done in \cite{bhtgr}
where we will show among other things that the chiral ring of the
$\cp{m}=G(m,m+1)$ model at level $k$ is the classical cohomology ring of
the Grassmannian $G(m,m+k)$.

The functional form of the observables and the calculation of correlation
functions is greatly simplified by the result of section 4 that the
topological \KS\ model can be localized and abelianized to a (perturbed)
$T/T$ model. In particular, therefore, all we ever need to know are
the operators $O_{l}$ and $\bar{O}_{l}$ evaluated for diagonal matrices
$t=\mbox{diag}(t_{1},\ldots,t_{n+m})$. In that case, the operators
reduce to the elementary symmetric functions of the $t_{k}$,
\bea
&&O_{l}(t) = \sum_{1\leq i_{1}<\ldots<i_{l}\leq m} t_{i_{1}}\ldots t_{i_{l}}
\;\;,\label{obs15}\\
&&\bar{O}_{l}(t) =
\sum_{m+1\leq j_{1}<\ldots<j_{l}\leq m+n} (t_{j_{1}}\ldots t_{j_{l}})^{-1}
\;\;.\label{obs16}
\eea

In applications we will find it convenient to parametrize the
torus valued field $t$ as
\be
t = \exp i\F\;\;,\label{tpar1}
\ee
where
\be
\F = \sum_{i=1}^{n+m-1}\a^{i}\f_{i}\;\;,\label{tpar2}
\ee
the $\a^{i}$ being simple roots of $SU(n+m)$,
\be
\a^{i} = E_{i,i}-E_{i+1,i+1}\;\;,\;\;\;\;\;\;(E_{i,j})_{kl} = \d_{ik}\d_{jl}
\;\;.
\ee
The correspondence with the above is then
\bea
&&t_{1}=\ex{i\f_{1}}\non
&&t_{k}=\ex{i(\f_{k}-\f_{k-1})}\;\;,\;\;\;\;\;\;k=2,\ldots,m+n-1\non
&&t_{m+n} = \ex{-i\f_{m+n-1}}\;\;.
\eea

Let us consider some examples. In the $\cp{1}$ model there is one
and only one scalar operator, namely
\be
O_{1} = g_{11}\;\;.
\ee
In the localized theory this becomes
\be
O_{1} = \ex{i\f_{1}}\;\;.
\ee
In the $\cp{2}$ model, there are two scalar operators, namely
\bea
&&O_{1} = g_{11} + g_{22} \ra \ex{i\f_{1}} + \ex{i(\f_{2}-\f_{1})}\non
&&O_{2} = g_{11}g_{22}-g_{12}g_{21} \ra \ex{i\f_{2}}\;\;.
\eea
And quite generally one finds that in the $G(m,m+n)$ models the
operator $O_{m}= \det g^{(m)}$ reduces to
\be
O_{m} = \bar{O}_{n} \ra \ex{i\f_{m}}\;\;.
\ee
As a final example, we consider the simplest Grassmannian which is
not a projective space, namely $G(2,4)$. In that case one has three
operators. In the localized and abelianized theory they are
\bea
&&O_{1}=\ex{i\f_{1}} + \ex{i(\f_{2}-\f_{1})} \non
&&O_{2}= \ex{i\f_{2}}\non
&&\bar{O}_{1}= \ex{i\f_{3}}+ \ex{i(\f_{2}-\f_{3})}\;\;.
\eea

\subsection{Bosonic Observables for Classical Flag Manifolds $G/T$}

We will now discuss the bosonic invariants for the
\ksm s based on the K\"ahlerian flag manifolds $G/T$ where
$G$ is a classical group and $T$ a maximal torus of $G$.
The strategy is exactly as above, i.e.\ we are looking for functionals of
$g$ which are invariant under the supersymmetry  (\ref{obs1})
and the gauge transformation $g \rightarrow t^{-1}gt$,
where $\alpha (\oa)$ now takes values in
the positive (negative) root spaces. In the
following we shall sometimes loosely use 
the term root to refer also to the Lie
algebra element associated with the root. A general reference for the
(rudiments of) Lie algebra theory used in the following is \cite{btd}.

{\bf 1. $SU(n)/T$}

Recall that $\dim SU(n) = n^{2}-1$  and $\rk SU(n) = n-1$, $T=U(1)^{n-1}$,
so that, on the basis of the rough argument sketched in section 6.1,
we expect to find a total of $n-1$ bosonic invariants. We will
see that this expectation is indeed borne out.

Let us dispose of gauge invariance first. The group valued field $g$ is
represented by an $n\times n$ matrix in $SU(n)$. Consider now the
$k\times k$ block matrix
\be
g_{ij}^{(k)}=g_{ij},(1\leq i\leq k, 1\leq j\leq k, 1\leq k\leq n)\,. \label{M}
\ee
One can now easily show that  $\det g_{ij}^{(k)}$ is gauge invariant as
follows. For $SU(n)$, $h$ is a diagonal $n\times n$ matrix
\be
h=\mbox{diag}(z_{1},\ldots,z_{n})\label{sut}
\ee
with $z_{i}=e^{i \theta_{i}}$, such that $z_{1}\dots z_{n}=1$.

One sees that under $g\rightarrow h^{-1}g h$ the matrix $g^{(k)}$
transforms as
\be
g^{(k)}\rightarrow (h^{-1})^{(k)}g^{(k)} h^{(k)}\;\;,
\ee
where
$h^{(k)}$ and $(h^{-1})^{(k)}$ are the upper left hand $k\times k$ blocks of
$h$ and $h^{-1}$, respectively. Also because of the diagonal nature of $h$ it
is easy to see that $(h^{-1})^{(k)}=(h^{(k)})^{-1}$ and hence that
$\det g^{(k)}$ is gauge invariant (as are other class functions of $g^{(k)}$,
of course, but these will not lead to BRST invariants).

Now let us check the first part of the supersymmetry transformation
(\ref{obs1}), namely
\be
\delta g_{ij}=g_{im}\alpha_{mj}\,.
\ee
For $SU(n)$ the positive roots can be taken to be
strictly upper triangular, i.e. $\alpha_{mj}=0$ for $m\geq j$.
A natural basis for the positive root spaces is
\be
\alpha^{[ij]}=E_{i,j}\;\;,,\;\;\;\;\;\; 1\leq i < j \leq n\;\;,
\ee
with $(E_{i,j})_{kl}=\d_{ik}\d_{jl}$, associated with the positive root
$\a^{i}+\ldots \a^{j-1}$, where $\a^{i}$ are the simple roots, $\a^{i}=
E_{i,i}-E_{i+1.i+1}$.
(Similarly the matrices belonging to the negative roots $\bar{\alpha}$
are strictly lower triangular matrices.) We note that for the elements
of the submatrix $g^{(k)}$ the supersymmetry transformation is restricted to:
\be
\delta g_{ij}^{(k)}= g_{im}\alpha_{mj}, 1\leq i\leq k, 1\leq j\leq k\,.
\ee
Since $\alpha$ is strictly upper triangular the only contributions to
this sum will come for $m<j$. Hence
the elements of $g$ that appear in the sum live only in the submatrix
$g^{(k)}$ itself. Therefore one may as well restrict ones attention to the
part of $\alpha$ living in the $k\times k$ block  $\alpha^{(k)}$,
\be
\delta g_{ij}^{(k)}= g_{im}^{(k)}\alpha^{(k)}_{mj}\;\;.
\ee
Then
\bea
\delta\det g^{(k)}&=&\det g^{(k)} Tr \left((g^{(k)})^{-1}\delta
g^{(k)}\right)\\
&=& \det g^{(k)} Tr\alpha^{(k)}=0
\eea
since $Tr\alpha^{(k)}=0$, $\alpha^{(k)}$ being strictly upper triangular.
 The same argument applies to
the other half of the supersymmetry transformation $\delta
g=\bar{\alpha} g$ where now the transformation of the matrices $g^{(k)}$ is
affected by multiplying on the left by  strictly lower triangular
matrices of the same dimension. Therefore all the
$\det g^{(k)},\;k=1,\ldots,n$, are gauge invariant and
supersymmetric. This gives rise to $n-1$ observables as one observable,
$\det g^{(n)}=\det g=1$, is trivial. Therefore
the number of scalar observables is equal to the rank of $SU(n)$.

Using the parametrization (\ref{tpar1},\ref{tpar2}) of the torus fields,
these observables take on the simple form
\be
\det g^{(k)}= \ex{i\phi_{k}}
\ee
in the localized $T/T$ theory.

{\bf 2. $SP(n)/T$}

First some facts about $SP(n)$. These are $2n\times 2n$ matrices $A$
satisfying $A^{T}\,J\,A=J$, where $J$ is the matrix
\be
J=\left[\begin{array}{cc}
0 &   -I      \\
I &    0
\end{array}
\right]\,,\label{app8}
\ee
with $I$ the identity matrix in $GL(n,C)$. The Lie algebra of $SP(n)$
consists of complex $2n\times 2n$ matrices
\be
{\cal A}=\left[\begin{array}{cc}
U &   W      \\
V &  -U^{T}
\end{array}
\right]\label{liesp}
\ee
where $V^{T}=V$ and $W^{T}=W$.

The dimension  of the group is
$n(2n+1)$ and the rank is $n$, so that the number of roots is $2n^{2}$.
 The torus is represented by diagonal
$2n\times 2n$ matrices
\be
h=\mbox{diag}(z_{1},\ldots,z_{n},\bar{z}_{1},\ldots,\bar{z}_{n})\,,
\label{sptorus}
\ee
with $z_{i}=e^{i\theta_{i}}$.

Again let us consider gauge invariance first. Let us write $g$ as the
$2n\times 2n$ matrix
\be
g=\left[\begin{array}{cc}
M &   N      \\
P &   Q
\end{array}
\right]\,.\label{gsp}
\ee
where $M,N,P$ and $Q$ are $n\times n$ matrices.
We see that under $g\rightarrow h^{-1}gh$, the $n\times n$ matrix $M$
transforms as
\be
M^{'}=\mbox{diag}(\bar{z}_{1},\ldots,\bar{z}_{n})
      M\mbox{diag}(z_{1},\ldots,z_{n})\;\;.
%\left[\begin{array}{cccc}
%\bar{z}_{1} &     &     &         \\
%	    &\cdot&     &         \\
%	    &     &\cdot&         \\
%	    &     &     &\bar{z}_{n}
%\end{array}
%\right] M
%\left[\begin{array}{cccc}
%z_{1} &     &     &         \\
%	    &\cdot&     &         \\
%	    &     &\cdot&         \\
%	    &     &     &z_{n}
%\end{array}
%\right]\,.
\ee
Thus following the same argument as in the $SU(n)/T$ case we see that
e.g.\ the determinant of every upper left hand $k\times k$
block of $M$,  $\det M^{(k)}, k=1,\ldots,n$, is gauge invariant.

For the supersymmetry we need to consider both the positive and
negative roots. We will only consider the positive roots explicitly as the
argument for the negative roots is then obvious. A basis for the $n^{2}$
elements of $\lk^{+}$ can be chosen as follows:

1. n matrices $A^{[i]}$, $1\leq i\leq n$ defined as
\be
A^{[i]}=E_{i,n+i}, 1\leq i\leq n\,.
\ee
 Referring to
(\ref{liesp}) this means that $U=V=0$ and $W$ is a diagonal matrix zero
everywhere except for $1$ at the $i,i$ position.

2. $\frac{n(n-1)}{2}$ matrices $B^{[ij]}$ defined as
\be
B^{[ij]}=E_{i,n+j}+E_{j,n+i}, (1\leq i\leq n, 1\leq j\leq, i<j)\,.
\ee
Again referring to (\ref{liesp}), this means that $U=V=0$ and $W$ is a
symmetric matrix with unit entries and zeroes along the diagonal.

3. $\frac{n(n-1)}{2}$ matrices $C^{[ij]}$ defined as
\be
C^{[ij]}=E_{ij}-E_{n+j,n+i}, (1\leq i\leq n, 1\leq j\leq, i<j)\,.
\ee
In this case $V=W=0$ and $U$ is a strictly upper triangular matrix with
unit entries.

Now we check for invariance under the supersymmetry transformation
$\delta g=g\alpha$ with $\alpha$ having values in the matrices
defined above. First of all we see immediately that for $\alpha$ in the
$A^{[i]}$ and $B^{[ij]}$ direction we have $\delta M=0$. For
$\delta g=gC^{[ij]}$ we see that either the $k\times k$ block, $M^{(k)}$, is
directly put to zero or it is multiplied by a $k\times k$ matrix with
entries only in the upper right hand corner and, therefore, by the same
arguments that we used in the $SU(n)$ case we have again $\det
M^{(k)}=0,\;k=1,\ldots,n$. Similar arguments apply to the other half of the
supersymmetry involving the negative roots. Thus again we come to the
conclusion that the number of observables is equal to the rank, in this
case $n$.

In this case, when we localize to the $T/T$ model, the
torus valued field, $g=t$, is parametrized as
\be
t=\mbox{diag}(\ex{i\phi_{1}},\ldots,\ex{i\phi_{n}},
\ex{-i\phi_{1}},\ldots,\ex{-i\phi_{n}})\,\label{soloc}
\ee
leading to the simple form of the observable
\be
\det M^{(k)}=\ex{i(\phi_{1}+\ldots +\phi_{k})}\,.\label{locobs}
\ee

{\bf 3. $SO(2n)/T$}

We have to do this example in a slightly unusual way to make use of what
we have learned from the previous examples. We would like to find the
torus and the Cartan subalgebra as diagonal matrices. $SO(2n)$ is
defined as $2n\times 2n$ matrices which satisfy
\be
A^{T}A=I
\ee
where $I$ is the unit matrix. The dimension of the group is $n(2n-1)$
while the rank is $n$ and hence the number of roots are $2n(n-1)$. Let us
now perform  a unitary transformation on A to yield an equivalent group of
matrices obeying a modified condition. Let us write
\be
A=UBU^{\dagger}\label{sotrans1}
\ee
so that
\be
A^{T}A=U^{\dagger T}B^{T}U^{T}UBU^{\dagger}=I\,.\label{sotrans2}
\ee
Put $K=U^{T}U$ to arrive at $B^{T}KB=K$. We will now work with this
group of matrices (it is easy to check that $A\rightarrow B$ is a
group homomorphism).  Writing
$B\simeq I+{\cal B} $ we see that the Lie algebra matrices $\cal B$ satisfy
\be
{\cal B}^{T}K+K{\cal B}=0\,.\label{lieso}
\ee

A convenient choice for the matrix $U$ is
\be
U=\frac{1}{\sqrt{2}}\left[\begin{array}{cc}
i_{n} &   -i_{n}      \\
-I_{n} &   -I_{n}
\end{array}
\right]\,,
\ee
where $i_{n}$ is  $i\times I_{n}$ with $I_{n}$ the identity $n\times n$
matrix. This leads to
\be
K=\frac{1}{\sqrt{2}}\left[\begin{array}{cc}
0 & I_{n}        \\
I_{n} &   0
\end{array}
\right]\,,
\ee

Writing $\cal B$ in terms of $n\times n$ matrices,
\be
{\cal B}=\left[\begin{array}{cc}
{\cal B}_{1} &   {\cal B}_{2}      \\
{\cal B}_{3} &   {\cal B}_{4}
\end{array}
\right]\,,
\ee
the condition (\ref{lieso}) gives
\be
{\cal B}_{1}=-{\cal B}_{4}^{T},\;{\cal B}_{2}=-{\cal B}_{2}^{T},\;
{\cal B}_{3}=-{\cal B}_{3}^{T}\,.\label{socon}
\ee
Here the torus is given by exactly the same matrices as in the case of
$SP(n)$, i.e. eq. (\ref{sptorus}). A basis for the $n(n-1)$ matrices of
$\lk^{+}$ belonging to the positive roots can be chosen as follows:

1. $\frac{n(n-1)}{2}$ matrices $B^{[ij]}$ defined as
\be
B^{[ij]}=E_{i,n+j}-E_{j,n+i}, (1\leq i\leq n, 1\leq j\leq, i<j)\,,
\ee
that is ${\cal B}_{1}={\cal B}_{3}={\cal B}_{4}=0$ with ${\cal B}_{2}$
an antisymmetric matrix with unit entries.

2.  $\frac{n(n-1)}{2}$ matrices $C^{[ij]}$ defined as
\be
C^{[ij]}=E_{ij}-E_{n+j,n+i}, (1\leq i\leq n, 1\leq j\leq, i<j)\,.
\ee
These are the same matrices as the $C$'s in $SP(n)$.

Now the rest of the argument follows exactly as in the $SP(n)/T$ case.
There are exactly $n$ observables (the rank of $SO(2n)$) given by $\det
B^{(k)}, (k=1,\ldots,n)$ where $B^{(k)}$ is the upper left $k\times k$
block of the group matrix $B$.

Upon localization, the torus valued field is parametrized in
exactly the same way, eq.(\ref{soloc}), as for the $SP(n)$ case and the
form of the observables here is the same
as in that case, eq.(\ref{locobs}).

{\bf 4. $SO(2n+1)/T$}

The dimension of the group $SO(2n+1)$ is $n(2n+1)$ while the rank is the
same as for $SO(2n)$, that is $n$. Hence the number of roots is $2n^{2}$.
For the odd dimensional orthogonal group we follow the same procedure
(\ref{sotrans1}-\ref{lieso}) as for the even dimensional case except that
we set
\be
U=\frac{1}{\sqrt{2}}\left[\begin{array}{ccc}
i_{n} &   -i_{n} & 0     \\
-I_{n} &   -I_{n} & 0     \\
  0    &    0    & \sqrt{2}
\end{array}
\right]\,,
\ee
so that
\be
K=\left[\begin{array}{ccc}
0_{n} &   I_{n} & 0     \\
I_{n} &   0_{n} & 0     \\
  0    &    0    & 1
\end{array}
\right]\,.
\ee
where as usual the subscript $n$ denotes a $n\times n$ matrix. The
corresponding $\cal B$ matrix can be written as
\be
{\cal B}=\left[\begin{array}{ccc}
{\cal B}_{1} &   {\cal B}_{2} & c_{1}     \\
{\cal B}_{3} &   {\cal B}_{4} & c_{2}     \\
   d_{1}     &   d_{2}        & b_{1}
\end{array}
\right]\,.
\ee
For the $2n\times 2n$ parts of the matrix, the conditions are the same
(\ref{socon}) as for the even dimensional algebra. The constraints on
the new matrices are
\be
b_{1}=0,\;\;\;c_{1}=-d_{2}^{T},\;\;\;c_{2}=-d_{1}^{T}\,.
\ee
The matrix for the torus $h$ is the same as for $SO(2n)$ except for a
$1$ added at the $2n+1,2n+1$ diagonal position,
\be
h=\mbox{diag}(z_{1},\ldots,z_{n},\bar{z}_{1},\dots,\bar{z}_{n},1)\,,
\label{sotorus}
\ee
For the basis of the $n^{2}$ Lie algebra elements belonging to the positive
roots we first choose $n(n-1)$ of them the same, $B^{[ij]}$ and $C^{[ij]}$,
as in the the $SO(2n)$ case except for a last column and row of zeros added.
The extra $n$ matrices of the basis for the new positive roots are
chosen as
\be
D^{[i]}=e_{i,2n+1}-e_{2n+1,n+i},\;1\leq i \leq n\,.
\ee
This means that ${\cal B}_{i}=0$ and $d_{1}=c_{2}=0$. This choice is
important because it ensures that $\delta B^{(n)}=0$ when $\alpha$ has values
in the direction of these roots, where as in the $SO(2n)$ case $B^{(n)}$ is
the upper left $n\times n$ block of $B\in SO(2n+1)$. Therefore  with this
choice
of basis it follows almost immediately that, as in the $SO(2n)$ case, we
have again exactly $n$ observables given by $\det B^{(k)},\; k=1,\ldots,n$.
Again, the localized observables have the same form as for the
$SP(n)$ and $SO(2n)$ cases.

\subsection{Fermionic Operators: Preliminary Remarks}

If there are fermionic zero modes, then in order to have non-vanishing
correlation functions they have to be soaked up (in a diffeomorphism
invariant way) by operators containing these fermionic zero modes.
Typically, in a cohomological field theory such operators can be constructed
from BRST invariant bosonic operators by what is known as `descent
equations'. Formally, they are a consequence of the BRST-exactness of
the energy momentum tensor (see e.g.\ \cite{rdlec}). In the present case
(where we are dealing with a topological field theory which is not quite of
the cohomological type) there are a few subtle differences. And while this
is only a minor issue and we don't want to overemphasize the differences,
we will nevertheless digress briefly to review the construction.

Before doing this, we want to point out that there is also
an alternative procedure for soaking up these zero modes which
bypasses the complications which arise when using the descent equations
and which essentially amounts to `dropping' the zero modes in a well-defined
way. First of all, we observe that as the $\b$-zero modes $\b^{0}$
do not appear in the action, one may define their BRST transformations
in any way one likes without spoiling the BRST symmetry of the action.
A natural choice is
\be
\d \b^{0}=  \d\ob^{0} = 0\;\;.
\ee
With this definition, gauge invariant combinations of the $\b$ and
$\bb$ zero modes like
\be
B^{l} = \Tr (\b^{0}\ob^{0})^{l}
\ee
are well-defined physical
operators that can be used to eliminate the fermionic zero modes
from the picture without changing anything else. In particular,
if one adopts this procedure one finds that there is a straightforward
correspondence between correlation functions in different topological
sectors. This has been discussed in terms of spectral flow in \cite{ns}.

Let us now review the descent equations.
The first of these expresses the fact that, given
a BRST invariant scalar operator $O^{(0)}$, one can find a one-form valued
operator $O^{(1)}$ satisfying
\be
dO^{(0)} = \d O^{(1)}\;\;.\label{obs17}
\ee
The main consequence of (\ref{obs17}) is that correlation functions of
scalar observables are independent of the points at which they have
been inserted - the hallmark of a topological field theory. What
is more important for us in the present context is that (\ref{obs17})
also implies that the integral of $O^{(1)}$ over a closed cycle $C$ is
also a BRST invariant operator,
\be
\d\oint_{C} O^{(1)} = \oint_{C} dO^{(0)} = 0 \;\;.\label{obs18}
\ee
The fact that the BRST cohomology class of this new operator only
depends on the homology class $[C]$ of this cycle is a consequence
of the second descent equation which says that one can find a two-form valued
operator $O^{(2)}$ satisfying
\be
d O^{(1)} = \d O^{(2)}\;\;.\label{obs19}
\ee
By the same argument as in (\ref{obs18}), (\ref{obs19}) implies that
the integral of $O^{(2)}$ over a closed two-cycle is a new BRST invariant
operator. In two dimensions the descent equations end here and one obtains
the one additional operator $\int_{\Sigma}O^{(2)}$.

In principle, therefore, one could now take any one of the bosonic
operators we constructed above and try to solve the equations
(\ref{obs17}) and (\ref{obs19}) by hand. In practice, however, this
is rather cumbersome and it is helpful to have a rather more conceptual
understanding of why the descent equations hold. Our starting point
is the fundamental equation of a topological field theory expressing
the fact that the energy-momentum tensor is BRST-exact,
\be
T_{\m\n} = \d G_{\m\n}\;\;.
\ee
Integrating this over a space-like hypersurface,  one obtains an equation
for the momentum $P_{\m}$ which in operator language this can be expressed
as
\be
d = \d G\label{ddg}
\ee
for some fermionic operator $G$.
Applying this to e.g.\ a scalar operator $O^{(0)}$ one recovers
(\ref{obs17}). The additional information one obtains is that
the one-form operator $O^{(1)}$ can be obtained from the scalar operator
by acting with the `vector supersymmetry' $G$,
\be
O^{(1)}= GO^{(0)}\;\;.
\ee
Likewise one finds
\be
O^{(2)}= GO^{(1)}\;\;,
\ee
etc. In a topological conformal field theory, (\ref{ddg}) can be written
more explicitly as
\be
\del_{z} = QG_{z}\;\;,\;\;\;\;\;\;\bar{\del}_{\zb}=\oQ\bar{G}_{\zb}
\;\;,\label{dqg}
\ee
where $Q$ ($\oQ$) and $G_{z}$ ($\bar{G}_{\zb}$) are the left (right)
moving BRST charges and supercurrents respectively.

\subsection{Fermionic Descendants for Grassmannian Models}

Let us now apply these considerations to \KS\ models and, more
specifically, to the observables of the Grassmannian models found
above. The first thing to note is that, as we have seen in section 5.1,
in the chosen Lagrangian realization of
the topological \KS\ model (as a supersymmetric gauged WZW model) the
energy-momentum tensor is only BRST-exact modulo the gauge field
equations of motion. This implies that a) operators depending on the
gauge field (like Wilson loops) will in general not lead to topological
correlation functions, and b) that we should expect to be able to solve
the descent equations only modulo the equations of motions of the gauge
fields.

Furthermore, the left- and right-moving BRST operators $Q$ and
$\oQ$ only commute up to the fermionic equations of motion. As a
consequence we will find that the descent-procedure described above,
while still a useful guideline, works somewhat differently in the
present context, in a sense reducing to the above only `on shell'. This
complicates the explicit determination of the one- and two-form
operators but is no fundamental obstacle to arriving at fermionic
descendants which have all the right properties inside correlation
functions.

With this in mind, let us no determine the relevant components of the
vector supersymmetry operator $G$. As we have seen, modulo the gauge field
equations of motion,
the $zz$-component of the energy-momentum tensor is $Q$-exact,
\be
T_{zz} = - Q\Tr \b_{z}^{-}J_{z}^{+}\;\;.
\ee
Choosing $\bar{z}$ as the time-coordinate in the left-moving sector,
the kinetic term of the action proportional to $D_{z}g^{-1} D_{\bar{z}}g$
implies that, as an operator, $D_{z} g^{-1}$ should be represented as
\be
D_{z}(g^{-1})_{kl} \rightarrow \frac{\d}{\d g_{lk}}\;\;.
\ee
Therefore, the counterpart of the operator $G_{z}$ in (\ref{dqg}) is
\be
G_{z} = \int dz\; (g\b_{z})_{li}\frac{\d}{\d g_{li}}
\equiv \Tr g\b\frac{\d}{\d g}\;\;.
\ee
Likewise for $\bar{G}_{\zb}$ one finds
\be
\bar{G}_{\bar{z}} = \int d\bar{z}\; \Tr (\ob g)\frac{\d}{\d g}\;\;.
\ee
{}From these equations one can read off directly the
candidates for the one-form descendants of the observables $O_{l}$ of the
Grassmannian \KS\ model,
\be
O_{l}^{(1)}=(G_{z}dz + \bar{G}_{\zb}d\zb)O_{l}\;\;.
\ee
It can be verified that these indeed satisfy (\ref{obs17}) and
(\ref{obs18}) if one uses the $A$-equations of motion.

For a number of reasons, however, these one-form operators appear not to be
of particular interest. For one, as there are always an equal number of
$\b$ and $\ob$ zero modes, the two-form operators $\sim \b\ob$ are
sufficient to soak these up. Moreover, the chiral ring is
determined by genus zero correlation functions and in genus
zero there are no non-trivial one-form operators. Finally,
it seems to be a general feature of these models that the
one-form operators are trivial inside correlation functions
as one is localizing onto BRST-invariant configurations.
In any case we will not consider further these one-form operators
here. The two-form operator, on the other hand, is of interest, and
it can be obtained from the scalar operator by operating on it
with the commutator of $G$ and $\bar{G}$,
\bea
O_{l}^{(2)}\equiv B_{l} &=& \trac{1}{2}[\bar{G},G]_{-}O_{l}(g)\non
&=&  \left[(\ob g)_{i_{1}k}(g\b)_{li_{2}}\frac{\d}{\d g_{i_{1}k}}
\frac{\d}{\d g_{li_{2}}} + (\ob g \b)_{i_{1}i_{2}}\frac{\d}{\d g_{i_{1}i_{2}}}
\right] O_{l}(g)\;\;.\label{ol2}
\eea
This time, when verifying (\ref{obs19})
one needs to use both the $A$ equations of
motion and $D_{\zb}\b_{z} = D_{z}\bb_{z} = 0$ which means that one is
essentially inserting only the zero modes of $\b$ and $\bb$. In the
path integral this is fine as long as none of the operators one is
considering depends on the Grassmann-odd scalars $\a$ and $\oa$, because
then integration over them will impose these equations.

As it stands, (\ref{ol2}) is valid quite generally and, in particular,
also for the observables of the $G/T$ models we discussed in section
6.3
(with the understanding that $\b$ then has components $\b_{ji}$ with
$i<j$ etc.).

We will of course ultimately be interested in doing calculations in the
localized and abelianized theory directly
(having kept only the zero modes of the fermionic fields).
As a preliminary
step one has to make sure that the operators we have constructed above
are also good operators in the perturbed $t$-dependent
theory we employed to establish localization. The point is that the
fermionic operators we discussed above have all the requisite properties
of an observable only modulo equations of motion and that these equations
of motion are $t$-dependent in the perturbed theory.
This issue as well as the calculation and (cohomological) interpretation
of correlation functions of these operators will be addressed elsewhere.

\subsection{A Comment on $\alpha$-Observables}

In the hermitian symmetric models, observables depending on the
Grassmann odd scalars $\a$ and $\oa$ are easy to come by and
permit one to consider correlation functions in the topological
sectors of the $U(1)$ gauge field in which there are zero modes
of these fields. The reason why this is particularly straightforward
in the hermitian symmetric models is that here $\a$ and $\oa$ are
already BRST invariant, $\d\a=\d\oa = 0$. To obtain gauge and BRST
invariant observables, it therefore suffices to take the invariants
(traces) of the matrices $\a\oa$ and $\oa\a$. We will show in
\cite{bhtgr} that the genus zero ring structure of these observables in the
hermitian symmetric $G/H$ level $k$ models is the classical cohomology
ring of $G/H$ - remarkably for all values of the level $k$.

\section{Non-Trivial Bundles}

In the preceding sections we have taken the point of view, even
though the
subgroup $F= H_{L}\times H_{R}$ of $G_{L}\times G_{R}$ is not necessarily
simply connected and the bundles in question may be non-trivial, that
results derived from the the WZW action (\ref{lraction}) are
nevertheless correct. The idea is that given some reasonable
definition of the WZW theory in general that both localization and
abelianization will apply and that ultimately one will be left with
the already well defined torus models of section 4. This section is
devoted to presenting one possible generalized WZW coset model that
fulfills our expectations.

Actually almost all coset models will require such an improved theory since
the question of the non-triviality of the
bundles involved has further complications. The extra difficulties
arise as the action of the group is the $\Ad$ action
so that the group that acts faithfully {\em is not} $F$ but is rather 
$F'= (H_{L} \times H_{R})/Z$ where $Z = H \cap Z(G)$. This means that
even though $F$ may be simply connected 
the gauge group need not be. Now $G$ bundles over a Riemann
surface are classified by $\pi_{1}(G)$. In particular
this means that one is often dealing with non-trivial $F'$ bundles. For
example consider a diagonal gauging of $SU(2)\times SU(2)$ by $SU(2)$,
then as the gauging is diagonal one is not really dealing with a
diagonal $F= SU(2)\times SU(2)$ bundle but rather a diagonal $F'= SO(3)
\times SO(3)$ bundle. In the following we will only consider connected
and simply connected $G$ (more general groups can be dealt with,
following \cite{gaw} but that would just serve to complicate
matters unnecessarily here).

The most immediate consequence of the appearance of non-trivial $F'$
bundles $P$ is that we are no longer dealing with maps
$g: \S \rightarrow  G$ but rather with sections of a non-trivial
bundle. At some point we will have to specify patching data to fix our
bundle. The fields on the various patches are then related by ``gauge
transformations'' on the overlaps. The metric part of the action and
the fermionic terms pose no quandary as they are manifestly gauge
invariant so that on the overlaps it makes no difference which
representative we take. We are left with the problem of making sense of
the gauged WZW term $\Gamma_{\S}(A;g)$ and, though not gauge
invariant, we would also like a
definition for the action of a wavefunction, $\Gamma_{\S}(A,B;g)$,
which keeps some of the
properties that are required in establishing the geometric nature of
the wavefunctions. Our task, then, is to give a definition of
$\Gamma_{\S}(A,B;g)$ in the case of non-trivial $H_{R,L}$ bundles so
that all the usual properties, such as the
Polyakov-Wiegmann identity, hold for the improved $\Gamma_{\S}(A,B;g)$,
which is denoted by $\Gamma_{\S ,P}(A,B;g)$ (the $P$ subscript stands for
the bundle). 

By considering Riemannn surfaces with boundary, Gawedzki \cite{gaw}
defines a line
bundle over the loop group (on the boundary $\partial \S = S^{1}$) with
the patching data being given by the exponential of the WZW action
subject to an equivalence relation. 
%The main reason for going through
%all of this is so that one can give a meaning to $\Gamma(g)$ for non-
%trivial bundles. 
The same construction, with the addition of a gauge
field was used by Hori \cite{hori} to define the coset models for non-trivial
bundles. We adopt and extend these constructions in order to define the action
for wavefunctions. In passing we note that the observables that we
have defined in section 6 do not require refinement as they are gauge
invariant and hence well defined for all bundles.

In the following subsections we will, very briskly, run through the
changes that one meets.

\subsection{$\Gamma_{\S ,P}(A,B;g)$}
That the definition of $\Gamma_{\S}(A;g)$ is inadequate when
non-trivial bundles are involved can be demonstrated with the use of a
simple example. We consider the $SU(2)/U(1)$ theory and let $g =
e^{i\f} \in
SU(2)$ lie in the torus. Such a configuration is, of course, very
special, however, we have seen in the body of the paper that such
configurations are precisely those that we would like to arrive at. For such
a configuration 
\be
\Gamma_{\S}(A;g) = -\frac{1}{2\pi}\int_{\S} A d\f \label{jump}
\ee
and the problem is manifest as the connection $A$ (and hence the
integrand) is not globally
defined  on $\S$.
A gauge invariant, and hence globally defined, functional
is obtained on adding a total derivative
\bea
\Gamma_{\S , P}(A;g)& =& \int_{\S} \f dA \nonumber \\
 &=& \Gamma_{\S}(A;g)-\frac{1}{2\pi} \int_{\S} d(A \f)
\eea

To give a general definition of $\Gamma_{\S , P}(A;g)$, and not
just for special configurations, requires a little more work even though
some, more or less, natural generalizations present themselves. Let us 
eliminate one such generalization of $\Gamma_{\S}(A,g)$. One can
express $\Gamma_{\S}(A,g)$ as
\be
\Gamma_{\S}(A;g) = \frac{1}{12\pi}\int_{M}\left( (g^{-1}d_{A}g)^{3} +
F_{A}(g^{-1}d_{A}g + d_{A}gg^{-1}) \right) \label{extension}
\ee
which is manifestly gauge invariant. For trivial bundles this is fine,
as we can extend a trivial bundle on $\S$ to a trivial bundle on
$M$. For non-trivial bundles this is not the case and will not do as a
definition of $\Gamma_{\S , P}(A;g)$. For example let $\S
= S^{2}$ and $M = D_{3}$ the unit three-disc. All bundles over $D_{3}$
are trivial (almost by definition), but by restricting ones attention
to the boundary of $D_{3}$ there cannot suddenly appear a non-trivial
bundle. Consequently, one cannot make sense of the right-hand side of
(\ref{extension}). 

To get around the problem of extending non-trivial bundles into
arbitrary three manifolds one takes a different tack. Rather than
dealing with sections of nontrivial bundles one gives a prescription
for producing maps from the sections of the bundle in question. The
WZW term for these maps is well defined, being the usual one. The bulk
of the work rests in establishing that nothing depends on the choices made
in producing the maps from the section. We now turn to that construction.

Decompose the Riemann surface $\S$ along a (homotopically trivial)
circle as $\S = D_{0} \# \S_{\infty}$. In order to define a
non-trivial $F'$ bundle it is enough to specify the overlap data with
a decomposition of this type.

On $\S$ we have $g$ as a section of a non-trivial
$F'$ bundle $P$ so that, with the above decomposition,
\be
g_{\infty}|_{\partial \S_{\infty}} = k^{-1}_{L}
 g_{0}|_{\partial D_{0}}  k_{R} . \label{overlap}
\ee
The patching data is given by the maps $(k_{L},k_{R}):
S^{1}\rightarrow (H'_{L},H'_{R}) \equiv ( H_{L},H_{R})/
 Z$. The right hand side of (\ref{overlap}) is well defined, when
thought of as a map into $G$, as the central element factors through.
Operationally this means that
$k_{L}$ and $k_{R}$ need not be periodic in $\theta$ (the angular
coordinate on the bounding circle). Rather, one requires that
\be
k_{L}(\theta + 2\pi) = zk_{L}(\theta) \;\;\;k_{R}(\theta + 2\pi) =
z k_{R}(\theta)
\ee
for some $z \in  Z$.

The aim now is to extend $g_{\infty}$ to a map on $\S$ and $g_{0}$ to a
map on the sphere $\CC\PP^{1}$. Let $\tilde{g}_{0}$ be a smooth map on 
$D_{0}$ to $G$ such that
\be
\tilde{g}_{0} \vee g_{\infty} : \S = \left( D_{0} \# \S_{\infty}\right) 
\rightarrow G 
\ee
is a smooth map. Likewise let $\tilde{g}_{\infty}$ be a smooth map on 
$D_{\infty}$ to $ G$ and chosen so that
\be
g_{0} \vee \tilde{g}_{\infty} : \CC\PP^{1}  = \left( D_{0} \# D_{\infty}
\right) \rightarrow  G 
\ee
is a smooth map. We will also need that $\tilde{g}_{\infty}$ be a
pointed map, that is at the origin $\{ 0\}$ of the disc
 $\tilde{g}_{\infty}(0) =I$ (or better still one may take the field
$\tilde{g}_{\infty} = I$ on some small open neighbourhood of $\{ 0
\}$)\footnote{One knows that it is always possible to find such maps.
Given a map from the boundary of the disc to $G$ one can construct a
map which is the identity at the origin of the disc in the following
way. The map at the boundary gives an $S^{1}$ in $G$, as $G$ is
simply connected one can always find a homotopy of that $S^{1}$ to
the identity element of $G$. Think of this homotopy as being a disc
with centre the identity element and boundary the $S^{1}$. Define a
map from $D_{\infty}$ to $G$ by the homotopy. This
homotopy will do as an example of $\tilde{g}_{\infty}$}. 
Notice that on the boundary circle the extensions satisfy
\be
\tilde{g}_{0}|_{\partial D_{0}}= k^{-1}_{L} \tilde{g}_{\infty}|_{
 \partial  D_{\infty}}k_{R} .
\ee

The gauge fields also satisfy an overlap equation; on $\S_{\infty}$ one
denotes the gauge fields by $A_{\infty}$ and $B_{\infty}$ while on the
disc $D_{0}$ one
denotes them by $A_{0}$ and $B_{0}$. On the overlap
\bea
A_{0} &=& k_{L}^{-1}A_{\infty}k_{L}+
k_{L}^{-1}dk_{L} \nonumber \\
B_{0} &=& k_{R}^{-1}B_{\infty}k_{R}+
k_{R}^{-1}dk_{R} .
\eea
We do not extend the gauge
fields. The gauged WZW action is perfectly local in the gauge fields
and so one does not need to improve that part of the story in this
construction.

Our definition of the gauge invariant Wess-Zumino-Witten term,
generalizing $\Gamma_{\S}(A,B;g)$ is
\bea
\Gamma_{\S ,P}(A,B;g) &=& \Gamma_{\S}(A_{\infty},B_{\infty};
\tilde{g}_{0} \vee g_{\infty}) + \Gamma_{{\Bbb C}{\Bbb P}^{1}}(A_{0},
B_{0}; g_{0} \vee \tilde{g}_{\infty}) \nonumber \\
& & -\Gamma_{{\Bbb C}{\Bbb P}^{1}}(\tilde{g}_{0} \vee
\hat{g}_{\infty}) +i {\cal
C}^{T}_{\S}(a_{L},b_{R};\tilde{g}_{\infty}) . 
\eea
The map $\hat{g}_{\infty}$ is defined by 
\be
\hat{g}_{\infty}= \tilde{g}_{\infty}^{\tilde{k}_{\infty}}=
\tilde{k}_{L \infty}^{-1}\tilde{g}_{\infty}\tilde{k}_{R \infty}
\ee
and the connections $a_{H_{L}}$ and $b_{H_{R}}$ are
\be
a_{L}= \tilde{k}_{L\infty}d\tilde{k}_{L\infty}^{-1} \;\;\;\; 
b_{R} = \tilde{k}_{R\infty}d\tilde{k}_{R\infty}^{-1}
\ee
where $\tilde{k}_{L,R \infty}$ is an extension of $k_{L,R}$ 
to the interior of
the disc minus the origin $D_{\infty}-\{ 0 \}$. 

For non-trivial $H$ bundles we are not be able to extend the $k_{(L,R)}$ to
the whole disc while remaining in $H_{(L,R)}$, for if we could we would
have produced a homotopy between the $k$ and a constant map and hence
be dealing with a trivial bundle\footnote{By allowing the $\tilde{k}$
fields to take values in $ G$ we can find extensions to the whole
disc since $\pi_{1}(G)=\pi_{2}(G)=0$. The subsequent
``trivialization'' of the bundle means that the gauge fields will take
values in $\lg$ even though one integrates only over a $\lh$'s
worth.}. Without such an extension it appears that $\hat{g}_{\infty}$
remains undefined, however, all is well as we have taken the extension
$\tilde{g}_{\infty}(0)= I$, as then at this point 
\be
\hat{g}_{\infty} = k_{L}^{-1}k_{R}
\ee
is globally defined (the non-trivial nature of the bundle comes from
either explicit $U(1)$ factors or from the modding through by $Z$).
Finally
\be
{\cal C}_{\S}(A,B;g) = \Gamma_{\S}(A,B;g)- \Gamma_{\S}(g)
\ee
is a local functional. $\tilde{g}_{0}\vee g_{\infty}$ is a smooth map
on $\S$,  $g_{0}\vee \tilde{g}_{\infty}$ is a smooth map on
$\CC\PP^{1}$ as is $\tilde{g}_{0}\vee \hat{g}_{\infty}$.
Thus every term in the generalized WZW functional is well
defined. When $H_{L}=H_{R}$ and $A=B$, our generalized gauged WZW term
devolves to Hori's \cite{hori} definition of a generalized gauged WZW
functional
$\Gamma_{\S ,P}(A;g)$
\be
\Gamma_{\S , P}(A;g) = \Gamma_{\S ,P}(A,A;g) , \label{hori}
\ee
with $b_{H}=a_{H}$. Let us also note that gauge transformations
$(h_{L},h_{R})$ satisfy
\be
h_{L \, 0} = k_{L}^{-1}h_{L \, \infty} k_{L}^{-1} , \;\;\; h_{L \, 0}
= k_{R}^{-1}h_{R \, \infty} k_{R}^{-1} 
\ee
on the overlap. The $Z$ identifications factor through, so that the
patching data for the gauge transformations is given completely by the
$(H_{L},H_{R})$ bundle. 

We now list, without proof, some of the most important properties of
this construction.

\noindent\underline{{\bf Properties:}}
\begin{enumerate}
\item $\Gamma_{\S ,P}(A,B;g)$ does not depend on the extension fields
$\tilde{g}_{0}$, $\tilde{g}_{\infty}$ and $\hat{g}_{\infty}$. \\
\item $\Gamma_{\S ,P}(A,B;g)$ does not depend on the extension maps
$\tilde{k}_{L}$ or $\tilde{k}_{R}$. \\
\item When the bundles are trivial $\Gamma_{\S ,P}$ reverts to
$\Gamma_{\S}$. \\
\item $\Gamma_{\S ,P}(A,B;g)$ satisfies the generalized
Polyakov-Wiegmann identity
\be
\Gamma_{\S ,P}(A^{h_{L}},B;h_{L}^{-1}g) = \Gamma_{\S ,P}(A,B;g) + \Gamma_{\S
,P}(h_{L}^{-1}) + \frac{i}{4\pi}\int_{\S}Adh_{L} h_{L}^{-1} \nonumber
\ee
where
\be
\Gamma_{\S ,P}(h_{L}^{-1}) = \Gamma_{\S ,P}(0;h_{L}^{-1})
\ee
with $h_{L\, 0}=k_{L}^{-1}h_{L \, \infty}k_{L}$ at the boundary.
There is a similar equation for $B$ gauge transformations.
\end{enumerate}

The first two assertions tell us that $\Gamma_{\S ,P}$ is well defined
and does not depend on any choices that we make. The third and fourth tell us
that it is a ``natural'' generalization of $\Gamma_{\S}$. The proofs
of these statements are straightforward but rather tedious (they
amount essentially to repeated use of the Polyakov Wiegmann identities).

\subsection{Action and Wavefunctions}
The action that we take to generalize to non-trivial bundles is
\be
S_{\S, P}(A,B;g)= -\frac{1}{8\pi}\int_{\S}g^{-1}d_{{\cal A}}g *
g^{-1}d_{{\cal A}}g  -i \Gamma_{\S ,P}(A,B;g)
\label{topact1} 
\ee
which, by virtue of property (4) above, satisfies the generalized
Polyakov-Wiegmann identities
\bea
S_{\S ,P}(A^{h_{L}},B;h_{L}^{-1}g) &=& S_{\S ,P}(A,B;g) +i\Gamma_{\S ,P}(h_{L})
-\frac{i}{4\pi} \int_{\S}A\, dh_{L} \,h_{L}^{-1} \nonumber \\
S_{\S ,P}(A,B^{h_{R}};gh_{R})& =& S_{\S ,P}(A,B;g) -i\Gamma_{\S ,P}(h_{R})
+\frac{i}{4\pi} \int_{\S}B\, dh_{R} h_{R}^{-1}  . \label{genpw}
\eea
The difference between the
gauge transformed action and the action depends neither on
the metric nor on the section $g$. Furthermore on setting
$H_{L}=H_{R}$ and taking $A=B$ we obtain a gauge invariant action
which is the action that we adopt when dealing with non-trivial bundles and
coincides with that introduced by Hori \cite{hori}.

Given the action (\ref{topact1}) we may form 
\be
\Psi_{P}(A,B;g) =  e^{-k S_{\S ,P}(A,B;g)}
\ee
and wavefunctions
\be
\Psi_{P}(A,B) = \int Dg e^{-k S_{\S ,P}(A,B;g)} = \int Dg \Psi_{P}(A,B;g)
\ee
where the path integral is over the space of sections of the bundle
$\{ P \}$, that is, over $(g_{0},g_{\infty})$ satisfying
(\ref{overlap}). When $P$ is trivial it is possible to interpret the
wavefunction as a holomorphic section of a product of line bundles
${\cal L}_{1}^{\otimes k}
\otimes {\cal L}_{2}^{\otimes (-k)} \rightarrow {\cal A} \otimes {\cal
B}$. However, in the present situation, it is better to think of the
space of gauge fields $(A,B)$ as one space ${\cal C}$ with gauge group
$(H_{L},H_{R})/Z$. On ${\cal C}$ there is still a symplectic two-form
given by (\ref{symp2f}) and one is given a single line bundle ${\cal L}
\rightarrow {\cal C}$ whose curvature agrees with the symplectic
two-form. One expects that $\Psi_{P}(A,B)$ is a holomorphic section
of ${\cal L}^{\otimes k}$. Under a gauge transformation one finds
\be
\Psi_{P}(A^{h_{L}},B^{h_{R}})= e^{ik\Phi_{P}}\Psi_{P}(A,B)
\ee
where
\be
\Phi_{P}= \Gamma_{\S ,P}(h_{R})-\Gamma_{\S ,P}(h_{L}) + \frac{1}{4\pi}
\int_{\S} \left( Adh_{L} h_{L}^{-1}-Bdh_{R} h_{R}^{-1}\right)
\ee

The generalized Polyakov-Wiegmann identities now guarantee
that we will obtain Ward identities analogous to those that hold in
the case of trivial bundles. If one takes infinitesimal gauge
transformations, $h_{L} = I + u_{L} +
\dots $ and $h_{R} = I + u_{R} + \dots$, the variation of the action
(\ref{topact1}) is
\be
\d S(A,B;g)= \frac{i}{4\pi} \int_{\S}\left( u_{L} dA -  u_{R}dB
\right) . \label{var}
\ee
The variation (\ref{var}) is exactly as we wrote it for `trivial'
bundles (\ref{vartriv}). 
Notice that while $\Gamma_{\S}(I + u + \dots) = 0 + \dots$, we have
instead that $\Gamma_{\S , P}(I + u_{L} + \dots ) = \frac{1}{4\pi}
\int_{\S}d(u_{L}A)$ (so
that in (\ref{var}) the exterior derivative is indeed in the correct spot).

$\Psi_{P}$ satisfies
\bea
\left( D^{A}_{\mu}\frac{\d }{\d A_{\mu}} + \frac{ik}{4\pi}\epsilon^{\mu
\nu} \partial_{\mu}A_{\nu}\right) \Psi_{P}(A,B) &=& 0 \nonumber \\
\left( D^{B}_{\mu}\frac{\d }{\d B_{\mu}} - \frac{ik}{4\pi}\epsilon^{\mu
\nu} \partial_{\mu}B_{\nu}\right) \Psi_{P}(A,B) &=& 0 \label{section}
\eea
with the variations and gauge fields appearing to be understood 
patch-wise. 
Likewise the following equations hold
\bea
\frac{D}{DA_{\bar{z}}} \Psi_{P}(A,B)&=& 0 \nonumber \\
\frac{D}{DB_{z}} \Psi_{P}(A,B)&=& 0 \label{hols}
\eea
providing one also understands them patch-wise. Put together (\ref{section})
and (\ref{hols}) give the more covariant
\bea
\left( D^{A}_{\mu}\frac{D }{D A_{\mu}} + \frac{ik}{4\pi}\epsilon^{\mu
\nu} F(A)_{\mu \nu}\right) \Psi_{P}(A,B) &=& 0 \nonumber \\
\left( D^{B}_{\mu}\frac{D }{D B_{\mu}} - \frac{ik}{4\pi}\epsilon^{\mu
\nu} F(B)_{\mu \nu}\right) \Psi_{P}(A,B) &=& 0 \label{sections}
\eea

\subsection{Factorization}
The wavefunctions have an important convolution
property when the gauge field $B$ takes its values in $\lg$. Fix the
Riemann surface $\S$. Let $g$
be a section of a $(H_{L}\times G)/Z$ bundle $P_{1}$ with connection
$(A,B)$ and let $h$ be a section of a $(G\times H_{R})/Z$ bundle
$P_{2}$ with connection $(B,C)$ (for this to make sense we require
$Z=H_{R}\cap Z(G)=H_{L}\cap Z(G)$). Then one finds
\be
\int DB \Psi_{P_{1}}(A,B;g)\Psi_{P_{2}}(B,C;h) =
\Psi_{P_{3}}(A,C;gh). \label{conv}
\ee
The integration over the gauge field $B$ certainly makes sense as the
product of the wavefunctions is invariant under $G$ gauge
transformations. Notice that the wavefunction on the right-hand side of
this expression involves a section $gh$ of a $(H_{L}\times H_{R})/Z$
bundle $P_{3}$ with connection $(A,C)$. 

Setting $H=H'$ and $A=C$ in (\ref{conv}) we reproduce the fact that a
$G/H$ model can be expressed as a norm of wavefunctions. Consequently
we are halfway to establishing holomorphic factorization. 

Even when $B$ does not live in all of $\lg$ the integral of a product
of wavefunction makes sense (is gauge invariant) and leads to a
respectable coset model. This remains true when we couple to
fermions. 

\subsection{Metric Variation}
The metric dependence of the coset models rests in $S_{0}(A,B;g)$
where
\be
S_{0}(A,B;g)= -\frac{1}{8\pi}\int_{\S}g^{-1}d_{(A,B)}g*g^{-1}d_{(A,B)}g.
\ee 
We
have not tampered with this part of the action, so that the metric
variation is as for the case of trivial bundles. Furthermore, as the
wavefunctions continue to satisfy (\ref{section}) we find that a
variation of the metric can be exactly cancelled by a second variation
of the $B$ field, that means that
\be
\nabla^{(1,0)}\Psi_{P}(A,B;\rho) =0. \label{ahol2}
\ee
As discussed above
\be
\int_{\S}\d \rho_{\bar{z}\bar{z}} \Tr
\frac{D}{DB_{\bar{z}}} \frac{D}{DB_{\bar{z}}}
\ee
must be understood patch-wise. The condition (\ref{ahol2}) is taken to
mean that $\Psi_{P}$ is an anti-holomorphic section of ${\cal W}$ for
non-trivial $P$. Thus, from this point of view, holomorphic
factorization has been established.

\section{Applications to the Kazama-Suzuki Coset Models}
The discussion of the previous section more or less covers the
wavefunctions for the supersymmetric and topological theories as well.
One keeps the fermionic part of the actions as given at the start of
this paper as they make sense as they stand. In this section we will
establish that replacing the bosonic part of the action $S_{\S}(A,B;g)$ by
$S_{\S,P}(A,B;g)$ does not spoil the supersymmetry. Consequently the
supersymmetric wavefunctions are taken care of. 

In order to explicitly solve the topological Kazama-Suzuki models we
had need of two main ingredients, localization and abelianization.
Here we will show that for 
non-trivial bundles these techniques still apply. The reason for this
is that working locally,
on patches, both techniques apply just as they did for the trivial
bundles, global information is regained by taking into account the
patching data $(k_{L},k_{R})=(k,k)$.

\subsection{Supersymmetry and Coupling to Topological Gravity}
The supersymmetry transformations (\ref{sup}) still leave the
action (with the bosonic action being $S_{P}$) invariant providing we
interpret them appropriately. On varying
the group element $g$ we understand that we are varying $g_{0}$ and
$g_{\infty}$ on their respective patches. This makes sense as, on the
right-hand side, the fermions (Grassmann variables) have the same
patching properties on the overlap. Now varying $g$ in (\ref{topact1})
leaves us with a local gauge invariant expression which is the same
that obtained on varying $S_{\S}(A,B;g)$. Hence the supersymmetry of
the various theories is guaranteed.

The same argument can be applied for the coupling to topological
gravity. The metric variation of the improved WZW action agrees (by
construction) with the variation of the usual WZW action. The crucial
point is that the gauge field variation of $S_{\S,P}(A,A;g)$ also
agrees with the gauge field variation of $S_{\S}(A;g)$. This is true
as in (\ref{topact1}) the gauge fields couple only to
$(g_{0},g_{\infty})$. Consequently in the action for the twisted
Kazama Suzuki model coupled to topological gravity one can replace
$S_{\S}(A,g)$ with $S_{\S ,P}(A;g)$ and retain the supersymmetry.

\subsection{Localization}
To implement localization in the case of trivial bundles we decomposed
the group element $g$ in  the $G/H$ Kazama-Suzuki models as
\be
g= h e^{i(\f^{+} + \f^{-})}
\ee
and scaled the coset fields $\f^{\pm} \rightarrow
t^{-1/2}\f^{\pm}$. To see that this strategy is justified even for
non-trivial bundles we work patch-wise. We let
\bea
g_{0}& =& h_{0}e^{i(\f^{+}_{0}+ \f^{-}_{0})} \nonumber \\
g_{\infty}& =& h_{\infty}e^{i(\f^{+}_{\infty}+ \f^{-}_{\infty})}
\eea
and on the overlap we have
\be
 h_{0}e^{i(\f^{+}_{0}+ \f^{-}_{0})}= k^{-1} h_{\infty}e^{i(\f^{+}_{\infty}+
\f^{-}_{\infty})} k
\ee

In the large $t$ limit after scaling we find that
\be
h_{0}\left( 1 + \frac{i}{\sqrt{t}} (\f^{+}_{0}+ \f^{-}_{0}) + \dots
\right) = k^{-1} h_{\infty}\left( 1 + \frac{i}{\sqrt{t}} (\f^{+}_{\infty}+
\f^{-}_{\infty}) + \dots \right)k
\ee
which splits into two equations
\be
h_{0} = k^{-1}h_{\infty}k \label{hoverlap}
\ee
and
\be
 (\f^{+}_{0}+ \f^{-}_{0})= k^{-1}(\f^{+}_{\infty}+
\f^{-}_{\infty})k
\ee
that is,
\be
 \f^{+}_{0} = k^{-1}\f^{+}_{\infty} k , \;\;\; \f^{-}_{0} =
k^{-1}\f^{-}_{\infty} k .  \label{cosetoverlap}
\ee
The first equation (\ref{hoverlap}) is the statement that $h$ is a
section of a non-trivial $H'$ bundle. The second equation
(\ref{cosetoverlap}) matches the conditions that are placed on the
Grassmann variables. Putting the two together we learn that on
localizing the ratio of determinants coming from the Grassmann
variables and those from $\f^{\pm}$ is, up to the standard anomaly,
unity and we are ultimately left with a $H/H'$ model for a non-trivial
$H'$ bundle.

\subsection{Abelianization}
We have now reduced the problem to calculating expectation values of
certain operators in a $H/H'$ model at some level. The $H$ valued
group fields (sections) are again defined locally with overlap data
specified. On each patch we can, up to the usual obstruction,
abelianize the group valued fields. Let the $H$ sections be denoted by
$h$. On $\S_{\infty}$ we have $h_{\infty}=
l^{-1}_{\infty}t_{\infty}l_{\infty}$, while on $D_{0}$, $h_{0}
=l^{-1}_{0}t_{0}l_{0}$, with $(t_{0}, t_{\infty}) \in T$. Abelianizing
patch-wise means that on the overlap we have
\be
t_{0}(\theta) = m_{0 \infty}^{-1}(\theta)t_{\infty}(\theta)m_{0
\infty}(\theta) , \;\; m_{0 \infty}(\theta)= l_{\infty}kl_{0}^{-1}.
\ee
We can organize things so that $m_{0\infty} \in T$ and, consequently,
that $t_{0} =
t_{\infty}$ on the overlap. For globally defined bundles $m_{0\infty}$
would be periodic in $\theta$, however, as $k$ is not periodic then
neither is $m_{0\infty}$. This feeds its way into the first Chern
class associated with the torus bundle. Put another way, the torus
field strength now encodes the information associated with the
original non-triviality of the $H'$ bundle as well as the
non-triviality of the `liberated' torus bundles. 

To see this explicitly for an $SU(n)/\Ad SU(n)$ bundle, let the
patching data be specified by
\be
k_{p}(\theta) = e^{ip/n \theta \lambda_{n-1}},
\ee
where $\lambda_{n-1}$ is the fundamental weight
$\frac{1}{n}(I_{n-1},1-n)$ ($I_{m}$ is the $ m\times m$ unit matrix),
so that $k_{p}(\theta + 2\pi) = e^{2\pi i p/n}k_{p}(\theta)$ with $p = 0,
\dots, n-1$. On abelianization the patching data for the $U(1)^{n-1}$
gauge fields is specified by the $m_{0\infty}$. The Chern classes are
then read off as
\bea
c_{1} &=& \frac{1}{2 \pi i}\oint m_{0\infty}^{-1}dm_{0\infty} \nonumber
\\
& =& \frac{p}{n}\lambda_{n-1} + \sum_{i=1}^{n-1}n^{i}\a_{i} \label{c}
\eea
where the $n^{i}$ are integers and the $\a_{i}$ are simple roots. One
arrives at the right-hand side in the following way. Since, $l_{0}$ and
$l_{\infty}$ are periodic we have
\be
m_{0\infty}(\theta + 2\pi) = e^{2\pi i p/n}m_{0\infty}(\theta)
\ee
and consequently
\be
m_{0\infty}(\theta) = k_{p}(\theta)e^{i\sum_{i=1}^{n-1}n^{i}\a_{i}\theta}.
\ee
The upshot is that the Chern class measures the non-triviality of the
original bundle (measured by $p$) and the non-triviality of the
liberated torus bundles (measured by the $n^{i}$). 

A detailed exposition of how to deduce the Chern classes for the
bundles that arise in the Grassmannian coset models is given in \cite{bhtgr}.

\subsection{Selection Rules}
Before gauging, the $G$ WZW action has a large global symmetry group,
including $G_{L} \times G_{R}$. Once the action is gauged, however,
the global invariance is considerably reduced. Indeed gauged WZW terms
appear in the bosonization of fermionic
systems precisely because they capture the the non-invariance of the
fermionic theory
under chiral transformations. Just as in the fermionic theory, global
chiral transformations in the gauged
WZW theories yield selection rules.

These selection rules, that govern the vanishing of correlation functions,
are easily derived at the level of the abelianized theory
\cite{ewcp,bhtgr}.  They are manifest at the abelian
level and correspond to shifts of the torus group valued
fields. However, one may
wonder how to arrive at them before abelianization. To derive a
selection rule we consider transformations on the fields $g
\rightarrow hg$ where $h$ is constant and commutes with the gauge
field $A$, $g$ and with $k$. The metric part of the
action is invariant under this change of variables and we must, therefore,
consider the effect on the WZW functional. We know that
\be
\Gamma_{\S , P}(A;hg) = \Gamma_{\S , P}(A;g) + \Gamma_{\S , P}(h)
\ee
for such an $h$. One would think that as $h$ is constant then it would
be possible to choose its extension into the bounding three manifold
to be constant so that $\Gamma_{\S , P}(h)=0$. This expectation is
naive, as in the definition of $\Gamma_{\S  , P}(A;hg)$ we meet a field
$\tilde{hg}_{\infty}= \tilde{h}_{\infty}\tilde{g}_{\infty}$ which
should satisfy $\tilde{hg}=I$ at the origin of the disc. Now we
have set $\tilde{g}_{\infty}=I$ at the origin so that $h$ cannot be
taken to be constant throughout. Consequently $\Gamma_{\S , P}(h)$ need
not be zero. 

We are allowed to consider any
extension of the element $h$ which is consistent with the condition
that $\tilde{h}_{\infty}(0)=I$ and that it is constant on
$\S_{\infty}$ and $D_{0}$ (taking the same value on the two
domains). On $D_{\infty}$ (thought of as a unit disc) we set
$\tilde{h}_{\infty}(r,\theta)=\tilde{h}(r)$ and demand that for some
$r_{0} \neq 0$, 
\be
\tilde{h}(r) = 0 , \;\;\; 0\leq r \leq r_{0} .
\ee
Furthermore, we demand that $\tilde{k}_{\infty}(r,\theta) =
\tilde{k}(\theta)$ and that the extensions $\tilde{h}$ and $\tilde{k}$
commute as matrices regardless of which points they are evaluated
at. With these simplifications plugging into the definition  of
$\Gamma_{\S , P}(h)$ we arrive at
\be
S_{\S, P}(A;hg)=S_{\S, P}(A;g) + \frac{\Tr}{2\pi}\oint_{S^{1}}\lambda
\tilde{k}^{-1} d \tilde{k},
\ee
where $h= e^{i\lambda}$.

As a simple example, if we consider the ``$SU(2)/U(1)$'' model we note
that $\oint k^{-1}dk = i\int_{\S}dA$ so that, in this case,
\be
S_{\S , P}(A;hg)= S_{\S, P}(A;g) + \frac{i}{2\pi}\int_{\S}\Tr \lambda dA
\ee
which agrees with the standard chiral anomaly and the result used in 
\cite{ewcp}.

\subsubsection*{Acknowledgements}

The research of M.B. was supported by a grant from the EC within the
framework of the HCM (Human Capital and Mobility) programme. We thank
K.S. Narain for numerous discussions and L. Alvarez-Gaum\'e for
an encouraging conversation. We would like to thank the referee for
supplying us with some references and suggestions which led us to
investigate the issues addressed in sections 7 and 8.

\rnc{\Large}{\normalsize}


\begin{thebibliography}{00}
\addcontentsline{toc}{section}{References}
\frenchspacing
\small
\addtolength{\itemsep}{-4pt}
\bibitem{ks}Y. Kazama, H. Suzuki, Nucl. Phys. B321 (1989) 232-268;
Phys. Lett.  B216 (1989) 112-116.
\bibitem{ewwzw} E. Witten, Nucl. Phys. B223 (1983) 422-432; Commun. Math.
Phys. 92 (1984) 455-472.
\bibitem{ewhf} E. Witten, Commun. Math. Phys. 144 (1992) 189-212.
\bibitem{lvw} W. Lerche, C. Vafa, N. Warner, Nucl. Phys. B324 (1989)
427-474.
\bibitem{gko} P. Goddard, A. Kent, D. Olive, Commun. Math. Phys. 103 (1986)
105-130.
\bibitem{schweigert} C. Schweigert, Commun. Math. Phys. 149 (1992) 425.
\bibitem{flmw}P. Fendley, W. Lerche, S. Mathur, N. Warner, Nucl. Phys.
B348 (1991) 66-88.
\bibitem{lw}W. Lerche, N. Warner, Nucl. Phys. B358 (1991) 571-599.
\bibitem{dg1}D. Gepner, Nucl. Phys. B322 (1989) 65-81.
\bibitem{dg2}D. Gepner, Commun. Math. Phys. 141 (1991) 381-411; Commun. Math.
Phys. 142 (1991) 433-491.
\bibitem{ht} S. Hosono, A. Tsuchiya, Commun. Math. Phys. 136 (1991)
451-486.
\bibitem{halpern} K. Bardacki, M.B. Halpern, Phys. Rev D3 (1971)
2493-2506. M.B. Halpern, Phys. Rev. D4 (1971) 2398-2401.
\bibitem{gkcoset} K. Gawedzki, A. Kupiainen, Nucl. Phys. B320 (1989) 625-668.
\bibitem{kscoset} D. Karabali, H. Schnitzer, Nucl. Phys. B329 (1990) 649-666.
\bibitem{schnitzer}H. Schnitzer, Nucl. Phys. B324 (1989) 412-426.
\bibitem{ewcp}E. Witten, Nucl. Phys. B371 (1992) 191-245.
\bibitem{nak} S. Nakatsu, Prog. Theor. Phys. 87 (1992) 795.
\bibitem{ns}T. Nakatsu, Y. Sugawara, Nucl. Phys. B405 (1993) 695-743;
hep-th/9304029.
\bibitem{heneg} M. Henningson, Nucl. Phys. B413 (1994) 73-83; hep-th/9307040.
\bibitem{henmir} M. Henningson, Nucl. Phys. B423 (1994) 631-638;
hep-th/9402122.
\bibitem{bhtgr} M. Blau, F. Hussain, G. Thompson. {\em
Grassmannian Topological Kazama-Suzuki Models and Cohomology}, ICTP and ENSLAPP
preprint IC/95/341, ENSLAPP-L-557/95, 44 p., {\em to appear}.
\bibitem{btlocrev} M. Blau, G. Thompson, J. Math. Phys. 36 (1995) 2192-2236;
hep-th/9501075.
\bibitem{btloc} M. Blau, G. Thompson, Nucl. Phys. B439 (1995) 367-394;
hep-th/9407042.
\bibitem{btver} M. Blau, G. Thompson, Nucl. Phys. B408 (1993) 345-390;
hep-th/9305010.
\bibitem{ewgr} E. Witten, {\em The Verlinde Algebra and the Cohomology of the
Grassmannian}, hep-th/9312104.
\bibitem{btdia} M. Blau, G. Thompson, Commun. Math. Phys. 172 (1995) 639-660;
hep-th/9412056.
\bibitem{hjs} J. Nunes, H. Schnitzer, {\em Field Strength Correlators for
Two Dimensional Yang-Mills Theories on Riemann Surfaces}, Brandeis preprint
BRX-TH-376, hep-th/9510154; {\em The Master Field for 2D QCD on the Sphere},
Brandeis preprint BRX-TH-385, hep-th/9510155.
\bibitem{gaw} K. Gawedzki, {\em Wess-Zumino-Witten Conformal Field
Theory}, Constructive Quantum Field Theory II, Erice (1988). 
\bibitem{hori} K. Hori, {\em Global Aspects of Gauged
 Wess-Zumino-Witten Models}, hep-th/9411134.
\bibitem{fp} A.N. Schellekens, Nucl. Phys. B366 (1991) 27-73. J.Fuchs,
A.N. Schellekens, C. Schweigert, Nucl. Phys. B461 (1996) 371-406;
hep-th/9509105. 
\bibitem{lerche} W. Lerche, Nucl. Phys. B434 (1995) 445-474; hep-th/9312188.
\bibitem{gawedzki} K. Gawedzki, {\em Constructive Conformal Field
                  Theory}, in {\em Functional Integration, Geometry and
              Strings},  eds. Z. Hava and J. Sobczyk (Birkhauser, 1989).
\bibitem{bdt} M. Blau, P. Degiovanni, G. Thompson, {\em unpublished}.
\bibitem{jfcs} J. Fuchs, C. Schweigert, Nucl. Phys. B411 (1994) 181-222;
hep-th/9304133.
\bibitem{cjetal} C. Johnson, Mod. Phys. Lett. A10 (1995) 549, hep-th/9409062;
P. Berglund, C. Johnson, S. Kachru, P.  Zaugg, {\em Heterotic Coset Models
and (0,2) String Vacua}, hep-th/9509170.
\bibitem{sstvp} P. Spindel, A. Sevrin, W. Troost, A. Van Proeyen,
Nucl. Phys. B308 (1988) 662; Nucl. Phys. B311 (1988) 465.
\bibitem{agew} A. Giveon, E. Witten, Phys. Lett. B332 (1994) 44-50;
hep-th/9494184.
\bibitem{ewsig} E. Witten, Commun. Math. Phys. 118 (1988) 411-449.
\bibitem{ey} T. Eguchi,  S.-K. Yang, Mod. Phys. Lett. A5 (1990) 1693-1701.
\bibitem{ehy} T. Eguchi, S. Hosono, S.-K. Yang, Commun. Math. Phys. 140
(1991) 159-168.
\bibitem{ewab} E. Witten, {\em Mirror Manifolds and Topological Field
Theory}, in {\em Essays on Mirror Manifolds} (ed. S.T. Yau), International
Press, Hong Kong (1992); hep-th/9112056.
\bibitem{btd} T. Br\"ocker, T. tom Dieck, {\em Representations of Compact
Lie Groups}, Springer, New York (1985).
\bibitem{rdlec} R. Dijkgraaf, E. Verlinde, H. Verlinde, Nucl. Phys. B352
(1991) 59-86.
\end{thebibliography}
\end{document}